\documentclass[sigconf]{acmart}
\AtBeginDocument{%
  }

\setcopyright{acmlicensed}
\copyrightyear{2018}
\acmYear{2018}
\acmDOI{XXXXXXX.XXXXXXX}
\acmConference[Conference acronym 'XX]{Make sure to enter the correct
  conference title from your rights confirmation email}{June 03--05,
  2018}{Woodstock, NY}
\acmISBN{978-1-4503-XXXX-X/2018/06}

\usepackage{xcolor}
\usepackage[table]{xcolor}
\definecolor{hlblue}{HTML}{F2ECF9}

\definecolor{deepurple}{HTML}{6A5ACD}
\definecolor{baselinegray}{HTML}{4D4D4D}  

\usepackage{multirow}

\usepackage{graphicx}  
\usepackage{subcaption}

\usepackage{amsthm}

\usepackage[ruled,vlined]{algorithm2e}

\usepackage{enumitem}
\setlist[itemize]{leftmargin=*}

\begin{document}

\title{BIPCL: Bilateral Intent-Enhanced Sequential Recommendation via Embedding Perturbation Contrastive Learning}

\author{Shanfan Zhang}
\affiliation{%
  \institution{School of software Engineering, \ Xi'an Jiaotong University}
  \city{Xi'an}
  \state{Shaanxi}
  \country{China}
}
\email{zhangsfxajtu@gmail.com}

\author{Yongyi Lin}
\affiliation{%
  \institution{School of Mathematics and Statistics, \ Xi’an Jiaotong University}
  \city{Xi'an}
  \country{China}}
\email{linyongyi@stu.xjtu.edu.cn}

\author{Yuan Rao}
\authornote{Corresponding author.}
\affiliation{%
  \institution{School of software Engineering, \ Xi'an Jiaotong University}
  \city{Xi'an}
  \state{Shaanxi}
  \country{China}
}
\email{raoyuan@mail.xjtu.edu.cn}

\renewcommand{\shortauthors}{Trovato et al.}

\begin{abstract}

  Accurately modeling users' evolving preferences from sequential interactions remains a central challenge in recommender systems. Recent studies emphasize the importance of capturing multiple latent intents underlying user behaviors. However, existing methods often fail to effectively exploit collective intent signals shared across users and items, leading to information isolation and limited robustness. Meanwhile, current contrastive learning approaches struggle to construct views that are both semantically consistent and sufficiently discriminative. In this work, we propose \emph{BIPCL}, an end-to-end \textbf{B}ilateral \textbf{I}ntent-enhanced, Embedding \textbf{P}erturbation-based \textbf{C}ontrastive \textbf{L}earning framework. \emph{BIPCL} explicitly integrates multi-intent signals into both item and sequence representations via a bilateral intent-enhancement mechanism. Specifically, shared intent prototypes on the user and item sides capture collective intent semantics distilled from behaviorally similar entities, which are subsequently integrated into representation learning. This design alleviates information isolation and improves robustness under sparse supervision. To construct effective contrastive views without disrupting temporal or structural dependencies, \emph{BIPCL} injects bounded, direction-aware perturbations directly into structural item embeddings. On this basis, \emph{BIPCL} further enforces multi-level contrastive alignment across interaction- and intent-level representations. Extensive experiments on benchmark datasets demonstrate that \emph{BIPCL} consistently outperforms state-of-the-art baselines, with ablation studies confirming the contribution of each component. All code and datasets are publicly available at \url{https://anonymous.4open.science/r/BIPCL-8E78/}.

\end{abstract}




\maketitle

\section{INTRODUCTION}

Sequential Recommendation (SR)~\cite{srec_survey,srec_survey_2,srec_survey_3,intent_works_4} is a core paradigm in modern recommender systems that models temporally ordered user interactions to predict future preferences. This temporal modeling enables SR to capture dynamic user behavior and has been extensively studied in both academia and industry.

Recent works~\cite{mgnm,intent_drift,intent_works_1,intent_works_2,intent_works_3} have emphasized the importance of intent learning, which aims to uncover latent user motivations that drive sequential behaviors. To model multiple user intents, representative methods employ capsule networks with dynamic routing or multi-head attention to decompose user behaviors into multiple intent vectors for target item matching. Early works such as \emph{MIND}~\cite{mind} and \emph{ComiRec}~\cite{comirec} pioneered this direction, while subsequent studies~\cite{pimirec,re4,umi,remi,simrec,dismir} incorporated temporal modeling, negative sampling, or regularization to mitigate routing instability and intent collapse. In parallel, some approaches incorporate intent awareness as auxiliary supervision to guide the learning of more discriminative sequence representations. Such methods typically leverage clustering or contrastive learning to identify latent intent prototypes and align sequence representations accordingly. Notable works such as \emph{ICLRec}~\cite{iclrec}, \emph{ICSRec}~\cite{icsrec}, and \emph{ELCRec}~\cite{elcrec} empirically demonstrate that intent-guided alignment improves both representation quality and robustness. Despite these advances, existing multi-intent models still face fundamental limitations.

\textbf{\emph{Limitations of Collective Intent Integration.}} 
Existing multi-intent recommenders primarily rely on intra-sequence disentanglement~\cite{mind,comirec,pimirec,re4,umi,remi,simrec,dismir}, employing mechanisms like dynamic routing to extract multiple intent vectors solely from an individual user's interaction history. While effective in capturing preference diversity, this design inherently constrains intent modeling to local behavioral contexts, limiting the ability to internalize collective behavioral regularities shared across the user population. Recent methods attempt to incorporate global signals via contrastive or clustering-based alignment~\cite{iclrec,icsrec,elcrec}. However, collective intents in these models mainly act as auxiliary supervision during training and affect representation learning only at the objective level. As a result, collaborative semantics are implicitly encoded in model parameters, but cannot be directly utilized for intent-specific reasoning or preference prediction at inference time. This non-explicit transfer of global behavioral priors leaves user representations heavily reliant on sparse and noisy single-sequence signals for preference estimation. These observations motivate models that preserve temporal dynamics while enabling more explicit and reliable utilization of collective intent information for robust generalization.

\emph{\textbf{Limitations of Contrastive View Construction.}} 
Contrastive learning is widely used in SR to enhance representation robustness and discrimination~\cite{cl_survey}. Its effectiveness critically depends on constructing contrastive views that balance diversity with semantic consistency. Existing methods~\cite{cl4srec,coserec,TiCoSeRec,elcrec,soft_cl,view_cl} typically generate contrastive views via discrete structural augmentations, including cropping, masking, or reordering item sequences. While inducing substantial variation, such operations may violate temporal order and disrupt long-range dependencies. Consequently, models can be encouraged to learn invariances that conflict with underlying behavioral semantics. Recent works~\cite{iclrec,icsrec,ioclrec} additionally construct contrastive views via clustering-based intent alignment. This design generally preserves semantic coherence by contrasting sequences against global latent intent representations, implicitly encouraging consistency among behaviorally related sequences. However, relying on aggregated intent prototypes may weaken instance-specific variation, leading to relatively homogeneous contrastive targets that provide limited discriminative signal. Overall, existing strategies struggle to generate contrastive views that are both semantically faithful and sufficiently discriminative.

\begin{figure}[t]
    \centering
    \includegraphics[width=\columnwidth]{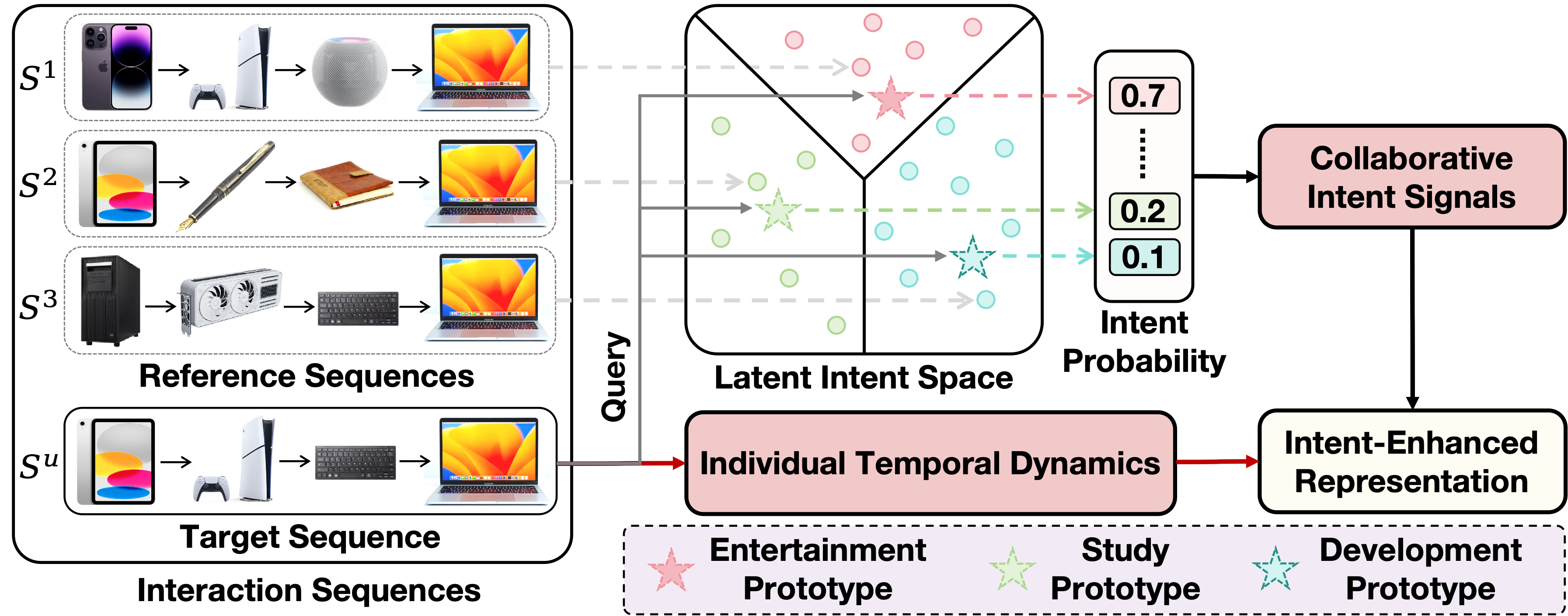}
    \caption{Collective Intent Integration in \emph{BIPCL}. Sequences ending with the same item can reflect different underlying intents. Capturing intent patterns shared across sequences allows sequence representations to incorporate intent-level semantics and support consistent recommendations.}
    \label{fig:intent_augs}
\end{figure}

\begin{figure}[t]
    \centering
    \includegraphics[width=\columnwidth]{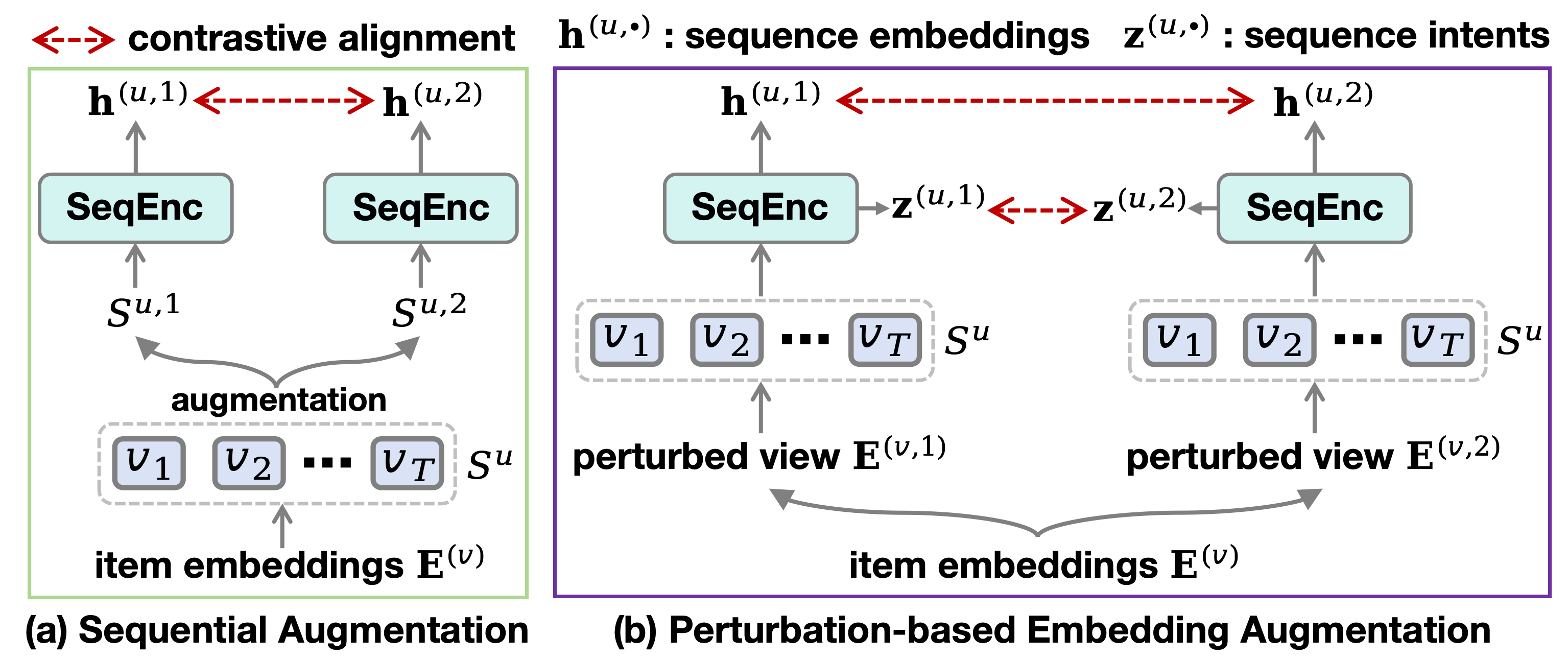}
    \caption{Comparison of contrastive view construction strategies in SR. SeqEnc denotes a generic sequence encoder over user history. \emph{BIPCL} (right) perturbs item embeddings to produce semantically consistent, multi-level contrastive views.}
    \label{fig:seq_augs}
\end{figure}

To address these limitations, we introduce \emph{BIPCL}, an end-to-end \textbf{B}ilateral \textbf{I}ntent-enhanced, \textbf{P}erturbation-based \textbf{C}ontrastive \textbf{L}earning framework. \emph{BIPCL} explicitly injects multi-intent signals into both item and temporal sequence representations via a bilateral intent-enhancement mechanism. On the item side, items are modeled as intrinsically multi-faceted entities through shared intent prototypes that capture collaborative semantic regularities. On the user side, \emph{BIPCL} decomposes user preferences into individual temporal dynamics and collective intent semantics. Individual interaction histories are used to extract personal preference patterns and align them with corresponding intent prototypes, thereby incorporating intent-level signals distilled from behaviorally related counterparts. By integrating collective intent semantics on both sides, \emph{BIPCL} mitigates the information isolation inherent in purely intra-sequence user modeling (Fig.~\ref{fig:intent_augs}) and co-occurrence-only item representations, enabling more robust preference estimation. All intent prototypes are learned jointly with the backbone model, facilitating intent-level sharing across users and items.

To construct high-quality contrastive views, \emph{BIPCL} injects bounded, direction-aware perturbations directly into structural item embeddings, instead of modifying the co-occurrence graph or input sequences that may distort item relations or temporal dependencies. The resulting perturbed embeddings are then propagated through the intent-enhancement modules to produce diverse yet semantically consistent views for both items and sequences. \emph{BIPCL} further enforces multi-level contrastive alignment across interaction- and intent-level representations, encouraging perturbation-invariant consistency at multiple semantic levels (Fig.~\ref{fig:seq_augs}). By jointly optimizing recommendation and contrastive objectives within a unified framework, \emph{BIPCL} achieves robust and discriminative intent-aware representations. Our key contributions are:
\begin{itemize}
    \item We introduce \emph{BIPCL}, an end-to-end bilateral intent-enhanced SR framework that explicitly incorporates collective intent signals into both user sequence and item representations.
    
    \item We design an embedding perturbation-based, multi-level contrastive learning paradigm that yields semantically faithful yet discriminative views with consistent representation alignment.
    
    \item Extensive experiments on multiple benchmarks showing that \emph{BIPCL} consistently outperforms state-of-the-art baselines, with ablations validating the effectiveness of each key component.
\end{itemize}

\section{METHODOLOGY}
We present the proposed \emph{BIPCL} framework. Sec.~\ref{sec:related_work} reviews the related work, and Sec.~\ref{procedure_code} summarizes the overall training procedure.

\subsection{Problem Definition}

Let $U$ and $V$ denote the sets of users and items, respectively. For each user $u \in U$, we observe an interaction sequence $\mathcal{S}^{u} = \left \{ s^{u}_{1},s^{u}_{2},\dots,s^{u}_{\left | \mathcal{S}^{u} \right | } \right \} $, where $s^{u}_{t} \in V$ denotes the item interacted at position $t$, and $|\mathcal{S}^{u}|$ is the sequence length. The objective of SR is to predict the item that user $u$ will interact with at the next position, formulated as:
\begin{equation}
    \mathrm{arg}\max_{v \in V} P\left ( s^{u}_{\left | \mathcal{S}^{u} \right | +  1}= v \mid \mathcal{S}^{u} \right )  
\end{equation}

\subsection{Bilateral Intent-Enhanced Structural Recommendation}

\subsubsection{\textbf{High-Order Structural Item Encoding}}

To capture collective item transition patterns beyond individual user sequences, we construct a global item co-occurrence graph $\textbf{A}\in \mathbb{R}^{\left | V \right |  \times \left | V \right | }$. For each user sequence $\mathcal{S}^{u}$, we accumulate co-occurrence increment between items $v_{i}$ and $v_{j}$ according to their positional distance $d_{ij}$:
\begin{equation}
    \textbf{A}_{ij} \leftarrow \textbf{A}_{ij} + \max(0, \delta - d_{ij})
\end{equation}
where $\delta$ controls the effective range of local transitions, assigning higher weights to closer item pairs~\cite{simrec}. We further set $\mathbf{A}_{ii}=1$ to preserve self-connections and apply row normalization to obtain the final graph. 
Given the initial item embedding matrix $\mathbf{E}^{\left ( v \right ) } \in \mathbb{R}^{\left | V \right |\times d } $, we perform lightweight graph propagation to incorporate high-order structural context:
\begin{equation}
    \mathbf{R}^{l} = \mathbf{A} \, \mathbf{R}^{l-1}, \quad l = 1, \dots, L, \quad \mathbf{R}^{0} = \mathbf{E}^{(v)}
\end{equation}
After $L$ propagation steps, each item embedding integrates information from its multi-hop neighborhood, producing structurally informed representations $\mathbf{R}^L \in \mathbb{R}^{\left | V \right |\times d }$. $\mathbf{R}^L$ serves as a global structural prior for subsequent intent-aware item and sequence modeling.

\subsubsection{\textbf{Item-Side Intent-Enhanced Representation Learning}}
\label{item_encoder}

In practical recommendation scenarios, items often exhibit multiple latent semantic facets. To explicitly model such semantics, we introduce a prototype-based intent modeling module that injects shared intent signals into item representations.

Specifically, we maintain a learnable intent prototype matrix $\mathbf{P}^{(v)} \in \mathbb{R}^{d \times K}$, where $K$ denotes the number of item intents and each column corresponds to a distinct intent prototype. Given the structural item embeddings $\mathbf{R}^L$, we derive intent-aware item representations via soft assignment over prototypes. The softmax is applied along the intent dimension, allowing each item to attend to multiple prototypes:
\begin{equation}
    \mathbf{Z}^{(v)} = \mathrm{softmax}\left(\mathbf{R}^L \, \mathbf{P}^{(v)}\right)\,\mathbf{P}^{(v)\top}
\end{equation}

These intent prototypes act as explicit semantic anchors, enabling each item to be represented as a weighted composition of multiple latent intents. We then employ a gated fusion mechanism to adaptively integrate structural and intent-level information:
\begin{equation}
    \mathbf{G}^{(v)} = \sigma\left(\left[\mathbf{R}^L \, \| \, \mathbf{Z}^{(v)}\right] \mathbf{W}_g^{(v)} \right)
\end{equation}
\begin{equation}
    \label{eq:item_gate}
    \mathbf{H}^{(v)} = \mathbf{R}^L + \mathbf{G}^{(v)} \odot \mathbf{Z}^{(v)}
\end{equation}
where $\mathbf{W}_g^{(v)}\in \mathbb{R}^{2d\times d} $ is a learnable projection, $\sigma(\cdot)$ denotes the sigmoid function, $\|\,$ concatenation, and $\odot$ element-wise multiplication. The resulting embeddings $\mathbf{H}^{(v)}  \in \mathbb{R}^{ \left | V \right |  \times d}$ simultaneously encode both global structural context and disentangled intent semantics.

\subsubsection{\textbf{User-Side Intent-Enhanced Sequential Representation Learning}}
\label{seq_encoder}

User behavior sequences encode evolving latent intents over time. We extend intent-aware modeling to the sequential level by explicitly injecting intent semantics into temporal representations. Given a fixed-length interaction sequence $\mathcal{S}^u = \{ s_1^u, \dots, s_T^u \}$, we retrieve the corresponding structural item embeddings and encode temporal dependencies via a sequential encoder:
\begin{equation}
    \mathbf{E}^{(u)} = \mathrm{TransEnc}\left(\mathbf{R}^L_{s_1^u}, \dots, \mathbf{R}^L_{s_{T}^u}\right)
\end{equation}
where $\mathbf{E}^{(u)} \in \mathbb{R}^{T \times d}$ contains contextualized hidden states. To jointly capture recency-sensitive behavioral signals and long-term preference regularities, we aggregate the final-step hidden state with the mean-pooled sequence embedding:
\begin{equation}
    \mathbf{e}^{(u)} = \frac{1}{2}\mathbf{E}^{(u)}_{T} + \frac{1}{2T}\sum_{t=1}^{T}\mathbf{E}^{(u)}_t
\end{equation}
as supported by the ablation results in Sec.~\ref{abs_study} (\textbf{w/o Pooling}) and Sec.~\ref{sec:case_study}. 
We then softly align the sequence representation with a set of shared user intent prototypes $\mathbf{P}^{(u)} \in \mathbb{R}^{d \times K}$:
\begin{equation}
    \mathbf{z}^{(u)} = \mathrm{softmax}\left(\mathbf{e}^{(u)} \mathbf{P}^{(u)}\right)\mathbf{P}^{(u)\top}
\end{equation}

Finally, gated fusion integrates sequential dynamics with intent semantics, where $\mathbf{W}_g^{(u)}\in \mathbb{R}^{2d\times d} $ is a learnable projection:
\begin{equation}
    \mathbf{g}^{(u)} = \sigma\left(\left[\mathbf{e}^{(u)} \, \| \, \mathbf{z}^{(u)}\right] \mathbf{W}_g^{(u)} \right)
\end{equation}
\begin{equation}
    \label{eq:seq_gate}
    \mathbf{h}^{(u)} = \mathbf{e}^{(u)} + \mathbf{g}^{(u)} \odot \mathbf{z}^{(u)}
\end{equation}
yielding the final sequence representation $\mathbf{h}^{(u)}  \in \mathbb{R}^{1 \times d}$.

\subsubsection{\textbf{Recommendation Objective}}

Given the sequence embedding $\mathbf{h}^{(u)}$ of user $u$, the objective is to predict the next interacted item by measuring its compatibility with candidate items. Let $\mathbf{h}^{(v)}_{j} \in \mathbb{R}^{1 \times d}$ denote the embedding of item $v_j \in V$, corresponding to the $j$-th row of $\mathbf{H}^{(v)}$. The relevance score is defined as:
\begin{equation}
s(u,v_j) = \left ( \mathbf{h}^{(u)} \right ) ^\top \mathbf{h}^{(v)}_{j}
\end{equation}

We optimize the model using a sampled softmax objective~\cite{sampled_softmax}. For each user sequence, we construct a candidate set $\mathcal{C}_u = \{v^+\} \cup \mathcal{N}_u$, where $v^+$ denotes the ground-truth next item and $\mathcal{N}_u$ contains $n$ negative items uniformly sampled from the item set $V$. The training loss with temperature hyperparameter $\tau_{1}$ is formulated as follows:
\begin{equation}
\label{eq:rec_loss}
\mathcal{L}_{\text{rec}}
= -\sum_{u\in U} \log
\frac{\exp\left(s(u,v^+)/\tau_{1}\right)}
{\sum_{v_j \in \mathcal{C}_u}\exp\left(s(u,v_j)/\tau_{1}\right)}
\end{equation}

This objective enables unified optimization of the initial item embeddings, intent prototypes, soft assignment weights, and gated fusion modules, encouraging effective discrimination between the ground-truth next item and negative candidates.

\subsection{Multi-level Contrastive Learning}

\subsubsection{\textbf{Perturbation-based Embedding Augmentation}}

We construct contrastive views directly in the embedding space to support contrastive learning while preserving semantic consistency.

Existing contrastive methods typically adopt two augmentation strategies: (1) perturbing the item–item interaction graph via edge dropping or subgraph sampling, or (2) augmenting user behavior sequences through masking, cropping, or reordering operations. Such strategies may distort intrinsic item relationships or disrupt temporal dependencies. Inspired by~\cite{fgsm,simgcl}, \emph{BIPCL} performs perturbations directly to learned structural item embeddings $\mathbf{R}^{L}$, avoiding modifications to the underlying graph topology or sequence order. Formally, given the structural embeddings $\mathbf{R}^{L}$, we generate two perturbed views by injecting bounded, direction-aware noise:
\begin{equation}
\widetilde{\mathbf{R}}^{(k)} = \mathbf{R}^{L} + \varepsilon \cdot \mathrm{sign}(\mathbf{R}^{L}) \odot \mathrm{norm}(\boldsymbol{\xi}^{(k)}), 
\quad k \in \{1,2\}
\end{equation}
where $\boldsymbol{\xi}^{(k)} \sim \mathcal{N}(\mathbf{0}, \mathbf{I})$ is Gaussian noise, $\mathrm{norm}(\cdot)$ denotes row-wise $\ell_2$ normalization, and $\varepsilon$ controls the perturbation magnitude. This formulation introduces stochastic yet bounded variations around the original embedding geometry. The sign-aligned perturbation respects local embedding directions, reducing semantic drift while encouraging robustness. Sec.~\ref{abs_study} validates its effectiveness.

The two perturbed views $\widetilde{\mathbf{R}}^{(1)}$ and $\widetilde{\mathbf{R}}^{(2)}$ are fed into the same item- and sequence-level intent-aware modules (Secs.~\ref{item_encoder}, \ref{seq_encoder}) by simply replacing $\mathbf{R}^{L}$. Importantly, contrastive learning shares an identical set of parameters with the recommendation task. As a result, parallel forward passes over the two perturbed views produce paired representations across multiple semantic levels:
\begin{itemize}
    \item interaction-level item representations $\mathbf{H}^{(v,1)}$, $\mathbf{H}^{(v,2)}$;
    \item item intent representations $\mathbf{Z}^{(v,1)}$, $\mathbf{Z}^{(v,2)}$;
    \item interaction-level user sequence representations $\mathbf{h}^{(u,1)}$, $\mathbf{h}^{(u,2)}$;
    \item user intent representations $\mathbf{z}^{(u,1)}$, $\mathbf{z}^{(u,2)}$.
\end{itemize}

By operating at the embedding level, this augmentation preserves item co-occurrence structure and temporal order, while enabling semantically consistent and informative contrastive views.

\subsubsection{\textbf{Multi-level Contrastive Alignment across Perturbed Views}}

To promote perturbation-invariant consistency and encourage more uniformly distributed representations, we perform contrastive alignment across multiple semantic levels.

We adopt the InfoNCE objective~\cite{infonce} to align corresponding representations derived from two perturbed views, while contrasting them against other instances of the same-type within a mini-batch. Formally, for a semantic entity $a$ (e.g., an item, an intent, or a user sequence), let $\mathbf{a}^{(1)}$ and $\mathbf{a}^{(2)}$ denote its representations obtained from the two perturbed views. The contrastive loss is defined as:
\begin{equation}
\label{eq:infonce_loss}
\mathcal{L}_{\mathrm{NCE}}(\mathbf{a}^{(1)}, \mathbf{a}^{(2)})
= - \log
\frac{
\exp\left(
\mathrm{sim}(\mathbf{a}^{(1)}, \mathbf{a}^{(2)})/\tau_{2}
\right)
}{
\sum_{b \in \mathcal{B}}
\exp\left(
\mathrm{sim}(\mathbf{a}^{(1)}, \mathbf{b}^{(2)})/\tau_{2}
\right)
}
\end{equation}
where $\mathrm{sim}(\cdot,\cdot)$ denotes cosine similarity, $\tau_{2}$ is a temperature hyperparameter, and $\mathcal{B}$ denotes the set of entities of the same semantic type as $a$ within the mini-batch. Here, $(\mathbf{a}^{(1)}, \mathbf{a}^{(2)})$ forms a positive pair, while the remaining instances in $\mathcal{B}$ serve as implicit negatives.

The overall contrastive objective aggregates alignment losses across all semantic levels:
\begin{align}
\label{eq:cl_loss}
    \mathcal{L}_{\mathrm{CL}} &=
\mathcal{L}_{\mathrm{NCE}}\!\left(\mathbf{h}^{(u,1)}, \mathbf{h}^{(u,2)}\right)
+ \mathcal{L}_{\mathrm{NCE}}\!\left(\mathbf{h}^{(v,1)}_{j}, \mathbf{h}^{(v,2)}_{j}\right)  \nonumber \\
&+ \mathcal{L}_{\mathrm{NCE}}\!\left(\mathbf{z}^{(u,1)}, \mathbf{z}^{(u,2)}\right)
+ \mathcal{L}_{\mathrm{NCE}}\!\left(\mathbf{z}^{(v,1)}_{j}, \mathbf{z}^{(v,2)}_{j} \right)
\end{align}

Each contrastive term regularizes representations at a distinct semantic granularity, preventing over-alignment at one level from suppressing diversity at others. By jointly enforcing contrastive alignment across item-, intent-, and sequence-level representations, \emph{BIPCL} achieves perturbation-invariant learning while preserving discriminative structure in the latent space.

\subsection{Overall Training Objective and Optimization}

\emph{BIPCL} is trained with a unified objective that jointly optimizes the recommendation loss and the multi-level contrastive alignment loss. The overall training objective is formulated as
\begin{equation}
\mathcal{L}
=
\mathcal{L}_{\mathrm{rec}}
+
\lambda \mathcal{L}_{\mathrm{CL}}
\end{equation}
where $\lambda$ controls the trade-off between recommendation accuracy and contrastive regularization. This unified objective explicitly integrates shared user- and item-level intent semantics into representations in an end-to-end manner, strengthening both item and sequence modeling. Contrastive supervision is achieved via lightweight embedding-level perturbations that reuse shared modules, enabling effective multi-level alignment with minimal overhead.

\subsection{Model Analysis}

\subsubsection{Complexity Analysis}
\label{complexity_ana}

Consider mini-batch training, $B$ is the mini-batch size. Structural propagation over the item co-occurrence graph is implemented via sparse matrix multiplication, incurring $\mathcal{O}(L|\mathcal{E}|d)$ time and $\mathcal{O}(|\mathcal{E}| + |V|d)$ space complexity. $|\mathcal{E}|$ is the number of non-zero edges. Prototype-based item intent modeling adds $\mathcal{O}(|V|Kd)$ cost. For sequential modeling, transformer-based encoding dominates the computation with $\mathcal{O}(BT^{2}d)$ time complexity, while intent aggregation and gated fusion add $\mathcal{O}(BKd)$ overhead. Contrastive learning operates on embedding-level perturbations and applies batch-wise InfoNCE objectives, resulting in $\mathcal{O}(B^{2}d)$ complexity without requiring additional encoders. Overall, \emph{BIPCL}'s time complexity is dominated by $\mathcal{O}(L|\mathcal{E}|d + |V|Kd)$, with space $\mathcal{O}(|\mathcal{E}| + |V|d)$. Since $B,T \ll |V|,|\mathcal{E}|$, the sequential and contrastive components introduce only minor overhead. All components are jointly optimized end-to-end in a single stage, without alternating optimization or iterative clustering.

\subsubsection{\textbf{Enhanced Expressivity and Optimization Dynamics}}
\label{sec:optimization_dynamics}

We analyze how intent-gated fusion modulates gradient flow to enhance training stability in \emph{BIPCL}. Recall Eq.~\ref{eq:seq_gate}, where the backbone sequence embedding $\mathbf{e}^{(u)}$ is combined with intent-aware features $\mathbf{z}^{(u)}$ via the gating vector $\mathbf{g}^{(u)}$.  

\textbf{\emph{Gradient decomposition.}} By the chain rule,
\begin{equation}
\frac{\partial \mathcal{L}_{\mathrm{rec}}}{\partial \mathbf{e}^{(u)}}
=
\frac{\partial \mathcal{L}_{\mathrm{rec}}}{\partial \mathbf{h}^{(u)}}
\left(
\mathbf{I}
+
\frac{\partial (\mathbf{g}^{(u)} \odot \mathbf{z}^{(u)})}{\partial \mathbf{e}^{(u)}}
\right)
\end{equation}
The gradient contribution from gating is
\begin{equation}
\label{eq:grad_gating}
\frac{\partial (\mathbf{g}^{(u)} \odot \mathbf{z}^{(u)})}{\partial \mathbf{e}^{(u)}}
=
\mathrm{Diag}(\mathbf{g}^{(u)}) \, \mathbf{J}_{\mathbf{z}}
+
\mathrm{Diag}(\mathbf{z}^{(u)}) \, \mathbf{J}_{\mathbf{g}}
\end{equation}
where $\mathbf{J}_{\mathbf{z}}=\partial \mathbf{z}^{(u)}/\partial \mathbf{e}^{(u)}$ captures how the intent representation varies with the backbone features, and $\mathbf{J}_{\mathbf{g}}=\partial \mathbf{g}^{(u)}/\partial \mathbf{e}^{(u)}$ captures how the gate adapts to the backbone state.

\textbf{\emph{Implications.}} The residual term $\mathbf{I}$ preserves a direct gradient path from $\mathbf{h}^{(u)}$ to $\mathbf{e}^{(u)}$, stabilizing optimization even when intent estimates are noisy or saturated early in training. The gated Jacobian terms enable intent-conditioned gradient routing: $\mathrm{Diag}(\mathbf{g}^{(u)}) \mathbf{J}_{\mathbf{z}}$ propagates gradients into the intent pathway, while $\mathrm{Diag}(\mathbf{z}^{(u)}) \mathbf{J}_{\mathbf{g}}$ adapts fusion weights based on reconstruction signals. This facilitates data-dependent allocation of learning capacity between sequential dynamics and intent semantics, enhancing representational flexibility without hard assignment or iterative procedures. A symmetric analysis applies to item-side gated fusion (Eq.~\ref{eq:item_gate}).

\subsubsection{\textbf{Uniformity Induced by Contrastive Alignment}}
\label{sec:uniformity}
We analyze how the InfoNCE objective promotes uniformity in latent space, ensuring stable and discriminative representations across interaction- and intent- levels.

\textbf{\emph{Gradient-based interpretation.}}
Recall the InfoNCE loss in Eq.~\ref{eq:infonce_loss}. Let $s_{b}=\mathrm{sim}(\mathbf{a}^{(1)}, \mathbf{b}^{(2)})$ and denote the softmax weights
\begin{equation}
p_b = \frac{\exp(s_b/\tau_{2})}{\sum_{\mathbf{c}\in\mathcal{B}} \exp(s_c/\tau_{2})}
\end{equation}
A direct differentiation yields
\begin{equation}
\frac{\partial \mathcal{L}_{\mathrm{NCE}}}{\partial s_b}
=
\frac{1}{\tau_{2}}\left(p_b - \mathbb{I}[\mathbf{b}=\mathbf{a}]\right)
\end{equation}
where $\mathbb{I}[\cdot]$ is the indicator function and $\mathbf{b}=\mathbf{a}$ denotes the positive pair. In particular, for any negative instance $\mathbf{b}\neq\mathbf{a}$,
\begin{equation}
\frac{\partial \mathcal{L}_{\mathrm{NCE}}}{\partial s_b}
=
\frac{p_b}{\tau_{2}} > 0
\end{equation}
Therefore, minimizing $\mathcal{L}_{\mathrm{NCE}}$ decreases $s_b$ for negatives, i.e., it pushes different instances in the batch away from the anchor. Moreover, since $p_b$ increases monotonically with $s_b$, \emph{more similar negatives receive larger gradients}, resulting in stronger repulsive forces among highly concentrated representations. This mechanism encourages a well-dispersed (uniform) embedding distribution in practice.

\textbf{\emph{Implications.}} \emph{BIPCL} applies contrastive alignment to both intent and interaction-level representations. The resulting uniformity among interaction embeddings helps reduce over-concentration along a few popular directions in the latent space (e.g., hubness), which may improve coverage for less frequent items. For intent embeddings, uniformity discourages degenerate configurations in which all intents become indistinguishable, reducing redundancy among prototypes and stabilizing multi-intent learning.

\textbf{\emph{Corollary.}} If all intent embeddings collapse to a single point, then for any anchor intent embedding, all negatives exhibit nearly identical similarity ($s_b \approx s_c$), implying $p_b \approx 1/|\mathcal{B}|$ and yielding persistent positive gradients. Consequently, the InfoNCE objective assigns high loss to such collapsed states and drives optimization toward more dispersed intent representations, without requiring explicit orthogonality constraints on intent prototypes.

\section{EXPERIMENTS}

\subsection{Experimental Setting}

\subsubsection{Datasets and Evaluation Protocol}

We conduct experiments on five public benchmark datasets: Beauty and Yelp~\cite{simrec}, as well as RetailRocket, Gowalla, and Amazon Books~\cite{dismir}. Following prior work~\cite{simrec,dismir}, all user-item interactions are treated as implicit feedback. Users or items with fewer than five interactions are filtered out to ensure sufficient behavioral history, and the maximum sequence length $T$ is truncated to 20. The statistics of the processed datasets are summarized in Table~\ref{tab:dataset}.

For fair comparison, we follow the standard evaluation protocol widely adopted in previous studies~\cite{mind,comirec,remi}. User sequences are split into training, validation, and test sets with an 8:1:1 ratio at the user level. Within the validation and test splits, the first 80\% of each user sequence is used to construct the input sequence, while the remaining 20\% serves as the ground-truth prediction targets. The full sequences in the training set are used for model learning. We evaluate all methods using standard ranking metrics, including Recall@N, Hit Rate@N (HR@N), and Normalized Discounted Cumulative Gain@N (NDCG@N), with $N\in\{20,50\}$.

\subsubsection{Baselines}

We compare \emph{BIPCL} against a comprehensive set of state-of-the-art multi-intent baselines spanning diverse modeling paradigms. 
\emph{MIND}~\cite{mind} employs capsule networks with dynamic routing to extract multiple latent user intents, while \emph{ComiRec}~\cite{comirec} improves routing stability via self-attention-based mechanisms (SA variant). 
\emph{Re4}~\cite{re4} further regularizes routing dynamics through backward flow constraints to alleviate intent collapse. 
\emph{UMI}~\cite{umi} and \emph{REMI}~\cite{remi} adopt hard or importance-aware negative sampling strategy to strengthen intent discrimination. \emph{PIMIRec}~\cite{pimirec} jointly considers temporal dynamics and item interactivity to enhance multi-intent learning. \emph{SimRec}~\cite{simrec} simulates item attributes derived from co-occurrence statistics, whereas \emph{DisMIR}~\cite{dismir} explicitly addresses representation collapse via global item partitioning and contrastive disentanglement. \emph{FRec}~\cite{frec} explicitly models user fatigue through intent-aware similarity features and fatigue-guided representation learning. \emph{GPR4DUR}~\cite{gpr4dur} leverages density-based user representations for adaptive and uncertainty-aware multi-intent retrieval. \emph{COIN}~\cite{coin} tackles intent entanglement and data sparsity via Transformer-based intent routing combined with collaborative neighbor alignment. \emph{ICSRec}~\cite{icsrec} models user intentions by performing contrastive learning over cross subsequences, capturing both shared and context-dependent intents through coarse- and fine-grained supervision. Overall, these baselines collectively cover major challenges in multi-intent learning, including intent disentanglement, routing stability, negative sampling, temporal dynamics, representation collapse, and cross-subsequence supervision. They provide a comprehensive benchmark for evaluating \emph{BIPCL}.

\subsubsection{Implementation Details}

We implement \emph{BIPCL} using PyTorch. For fair comparison, all methods use the Adam optimizer with embedding dimension 64 and batch size 256. Following~\cite{dismir}, we sample $n=10$ negatives per sequence (50 for Amazon Books due to its larger item space). We instantiate $\mathrm{TransEnc}(\cdot)$ as a Transformer-based sequential encoder. For small datasets (Beauty, Yelp), we use a single self-attention block, while two blocks are employed for larger datasets. All configurations adopt 4 attention heads. The item co-occurrence window size is set to $\delta=5$ for all datasets except Beauty, where $\delta=3$ is used, and the propagation depth for structural item embeddings is $L=2$. The number of intent prototypes is $K=256$ for small datasets and $K=512$ for larger ones. The reconstruction and contrastive loss temperatures are set to $\tau_1=1.0$ and $\tau_2=0.2$, respectively. The embedding perturbation magnitude is $\varepsilon=0.1$, and the loss balancing coefficient is $\lambda=50$. For baseline models, we use the authors' original hyperparameters when available; otherwise, we tune them on validation sets.

\begin{table}[t]
  \caption{Statistical summary of the experimental datasets.}
  \label{tab:dataset}
  \begin{center}
  \resizebox{\columnwidth}{!}{
    \begin{tabular}{lcccr}
      \toprule
      Dataset  
      & \# Users & \# Items  & \# Actions & Sparsity\\
      \midrule
      Beauty\cite{simrec}  & 22,363  & 12,101  & 198,502  & 0.0734\%   \\
      Yelp\cite{simrec}    & 19,242  & 14,142  & 201,237  & 0.0740\%   \\
      Retail Rocket\cite{dismir}  & 33,708  & 81,635  & 356,840  & 0.0130\%  \\
      Gowalla\cite{dismir} & 165,506 & 174,605 & 2,061,264  & 0.0071\% \\
      Amazon Books\cite{dismir}   & 603,668 & 367,982 & 8,898,041  & 0.0040\% \\
      \bottomrule
    \end{tabular}
    }
  \end{center}
\end{table}

\begin{table*}[t]
  \caption{Overall performance comparison on five benchmark datasets.
Bold and underlined values indicate the best and second-best performance, respectively. R@N denotes Recall@N, and ND@N denotes NDCG@N.}
  \label{tab:compare_perform}
  \begin{center}
      \resizebox{\textwidth}{!}{
        \begin{tabular}{cccccccccccccccc}
          \toprule
          Dataset  & Metric
          & MIND     &  ComiRec 
          & Re4      & UMI      &  PIMIRec 
          & REMI     & SimRec   &  DisMIR
          & FRec     & GPR4DUR  &  COIN  & ICSRec
          & ours 
          & \emph{Improv.} \\
          \midrule

        \multirow{6}{*}{Beauty}
          & R@20
          & 0.0822  & 0.0650
          & 0.0741 & 0.0795 & 0.0731
          & 0.0779 
          & \underline{\textcolor{baselinegray}{0.1139}}
          & 0.1097
          & 0.0896 & 0.0855 & 0.0963  & 0.1105
          & \textbf{\textcolor{deepurple}{0.1277}}  
          & \cellcolor{hlblue} 12.12\% $\uparrow$ \\
          
          & R@50   
          & 0.1259 & 0.1102
          & 0.1142 & 0.1279 & 0.1101 
          & 0.1259 
          & \underline{\textcolor{baselinegray}{0.1810}}
          & 0.1798
          & 0.1507 & 0.1392 & 0.1564  & 0.1733
          & \textbf{\textcolor{deepurple}{0.1861}}  
          & \cellcolor{hlblue} 2.82\% $\uparrow$ \\
          
          & ND@20 
          & 0.0650 & 0.0415
          & 0.0539 & 0.0501 & 0.0477 
          & 0.0537 
          & \underline{\textcolor{baselinegray}{0.0758}}
          & 0.0730
          & 0.0615 & 0.0589 & 0.0676  & 0.0731
          & \textbf{\textcolor{deepurple}{0.0909}}  
          & \cellcolor{hlblue} 19.92\% $\uparrow$ \\
          
          & ND@50  
          & 0.0769 & 0.0554
          & 0.0665 & 0.0651 & 0.0631
          & 0.0687 
          & \underline{\textcolor{baselinegray}{0.0927}}
          & 0.0886
          & 0.0773 & 0.0714 & 0.0838 & 0.0893
          & \textbf{\textcolor{deepurple}{0.1052}}  
          & \cellcolor{hlblue} 13.48\% $\uparrow$ \\
          
          & HR@20   
          & 0.1390 & 0.1100
          & 0.1229 & 0.1265 & 0.1305 
          & 0.1346 
          & \underline{\textcolor{baselinegray}{0.1873}}
          & 0.1694
          & 0.1529 & 0.1332 & 0.1578 & 0.1815
          & \textbf{\textcolor{deepurple}{0.2074}}  
          & \cellcolor{hlblue} 10.73\% $\uparrow$ \\
          
          & HR@50  
          & 0.1989 & 0.1775
          & 0.1873 & 0.2003 & 0.2070 
          & 0.2088 
          & \underline{\textcolor{baselinegray}{0.2758}}
          & 0.2669
          & 0.2316 & 0.2062 & 0.2387 & 0.2677
          & \textbf{\textcolor{deepurple}{0.2839}}  
          & \cellcolor{hlblue} 2.94\% $\uparrow$ \\
          
        \midrule

    \multirow{6}{*}{Yelp}
          & R@20
          & 0.0624  & 0.0625 
          & 0.0598 & 0.0737  & 0.0842 
          & 0.0725 & 0.0921  
          &  \underline{\textcolor{baselinegray}{0.0992}}
          & 0.0767 & 0.0741 & 0.0827 & 0.0960
          & \textbf{\textcolor{deepurple}{0.1228}}  
          & \cellcolor{hlblue} 23.79\% $\uparrow$ \\
          
          & R@50   
          & 0.1373  & 0.1282
          & 0.1183 & 0.1408  & 0.1487 
          & 0.1328 & 0.1881  
          & \underline{\textcolor{baselinegray}{0.1935}}
          & 0.1452 & 0.1408 & 0.1528 & 0.1720
          & \textbf{\textcolor{deepurple}{0.2190}}  
          & \cellcolor{hlblue} 13.18\% $\uparrow$ \\
          
          & ND@20 
          & 0.0479  & 0.0547 
          & 0.0489 & 0.0571  & 0.0644 
          & 0.0638 & 0.0679  
          & 0.0740
          & 0.0591 & 0.0642 & 0.0674 
          & \underline{\textcolor{baselinegray}{0.0766}}
          & \textbf{\textcolor{deepurple}{0.0963}}  
          & \cellcolor{hlblue} 25.72\% $\uparrow$ \\
          
          & ND@50  
          & 0.0764  & 0.0787
          & 0.0695 & 0.0806  & 0.0861 
          & 0.0852 & 0.0984  
          & 0.1023
          & 0.0846 & 0.0868 & 0.0918 
          & \underline{\textcolor{baselinegray}{0.1027}}
          & \textbf{\textcolor{deepurple}{0.1274}}  
          & \cellcolor{hlblue} 24.05\% $\uparrow$ \\
          
          & HR@20   
          & 0.1319  & 0.1340 
          & 0.1210 & 0.1491  & 0.1668 
          & 0.1616 & 0.1839  
          & 0.1870
          & 0.1532 & 0.1647 & 0.1694 
          & \underline{\textcolor{baselinegray}{0.1938}}
          & \textbf{\textcolor{deepurple}{0.2343}}  
          & \cellcolor{hlblue} 20.90\% $\uparrow$ \\
          
          & HR@50  
          & 0.2655  & 0.2556
          & 0.2270 & 0.2644  & 0.2784 
          & 0.2732 & 0.3345  
          & \underline{\textcolor{baselinegray}{0.3414}} 
          & 0.2722 & 0.2806 & 0.2868 &  0.3247
          & \textbf{\textcolor{deepurple}{0.3943}}  
          & \cellcolor{hlblue} 15.50\% $\uparrow$ \\

        \midrule

    \multirow{6}{*}{\shortstack{Retail\\Rocket}}
          & R@20
          & 0.1415  & 0.1035 
          & 0.1397 & 0.1519  & 0.1828 
          & 0.2129 & 0.2259  
          & \underline{\textcolor{baselinegray}{0.2385}}
          & 0.1735 & 0.2046 & 0.2189 & 0.2298
          & \textbf{\textcolor{deepurple}{0.2580}}  
          & \cellcolor{hlblue} 8.18\% $\uparrow$ \\
          
          & R@50   
          & 0.2148  & 0.1666 
          & 0.2194 & 0.2423  & 0.2811
          & 0.3160 & 0.3433
          & \underline{\textcolor{baselinegray}{0.3447}} 
          & 0.2794 & 0.3107 & 0.3356 &  0.3419
          & \textbf{\textcolor{deepurple}{0.3813}}  
          & \cellcolor{hlblue} 10.62\% $\uparrow$ \\
          
          & ND@20 
          & 0.0804  & 0.0609 
          & 0.0785 & 0.0875  & 0.1025 
          & 0.1198 & 0.1210 
          & \underline{\textcolor{baselinegray}{0.1330}} 
          & 0.0997 & 0.1133 & 0.1206 & 0.1302
          & \textbf{\textcolor{deepurple}{0.1649}}  
          & \cellcolor{hlblue} 23.98\% $\uparrow$ \\
          
          & ND@50  
          & 0.0880  & 0.0684
          & 0.0884 & 0.0974  & 0.1093 
          & 0.1281 
          & 0.1378
          & 0.1360 
          & 0.1022 & 0.1213 & 0.1322 
          & \underline{\textcolor{baselinegray}{0.1496}}
          & \textbf{\textcolor{deepurple}{0.1918}}  
          & \cellcolor{hlblue} 28.21\% $\uparrow$ \\
          
          & HR@20   
          & 0.2195  & 0.1602 
          & 0.2103 & 0.2364  & 0.2764 
          & 0.3183 & 0.3349
          & \underline{\textcolor{baselinegray}{0.3524}} 
          & 0.2616 & 0.3052 & 0.3195 &  0.3453
          & \textbf{\textcolor{deepurple}{0.3788}}  
          & \cellcolor{hlblue} 7.49\% $\uparrow$ \\
          
          & HR@50  
          & 0.3183  & 0.2501 
          & 0.3174 & 0.3574  & 0.3969
          & 0.4515 & 0.4758
          & \underline{\textcolor{baselinegray}{0.4815}} 
          & 0.3878 & 0.4365 & 0.4616 &  0.4794
          & \textbf{\textcolor{deepurple}{0.5215}}  
          & \cellcolor{hlblue} 8.31\% $\uparrow$ \\

        \midrule
    
    \multirow{6}{*}{Gowalla}
          & R@20
          & 0.0923  & 0.0623
          & 0.0843 & 0.0961  & 0.1193
          & 0.1300 & 0.1211  
          & \underline{\textcolor{baselinegray}{0.1384}} 
          & 0.1045 & 0.1122 & 0.1273 & 0.1286
          & \textbf{\textcolor{deepurple}{0.1506}}  
          & \cellcolor{hlblue} 8.82\% $\uparrow$ \\
          
          & R@50   
          & 0.1521  & 0.1181
          & 0.1396 & 0.1642  & 0.1951 
          & 0.2109 & 0.2056
          & \underline{\textcolor{baselinegray}{0.2169}} 
          & 0.1796 & 0.1939 & 0.2094 &  0.2125
          & \textbf{\textcolor{deepurple}{0.2336}}  
          & \cellcolor{hlblue} 7.70\% $\uparrow$ \\
          
          & ND@20 
          & 0.1336  & 0.0955 
          & 0.1287 & 0.1391  & 0.1603
          & 0.1715 & 0.1696
          & \underline{\textcolor{baselinegray}{0.1855}} 
          & 0.1457 & 0.1634 & 0.1742 & 0.1833
          & \textbf{\textcolor{deepurple}{0.2164}}  
          & \cellcolor{hlblue} 16.66\% $\uparrow$ \\
          
          & ND@50  
          & 0.1450  & 0.1199 
          & 0.1410 & 0.1505  & 0.1660 
          & 0.1792 & 0.1825
          & 0.1849
          & 0.1572 & 0.1718 & 0.1826 
          & \underline{\textcolor{baselinegray}{0.1960}}
          & \textbf{\textcolor{deepurple}{0.2377}}  
          & \cellcolor{hlblue} 21.28\% $\uparrow$ \\
          
          & HR@20   
          & 0.3122  & 0.2281 
          & 0.3104 & 0.3314  & 0.3843 
          & 0.4021 & 0.4012
          & \underline{\textcolor{baselinegray}{0.4312}} 
          & 0.3387 & 0.3914 & 0.4087 &  0.4174
          & \textbf{\textcolor{deepurple}{0.4485}}  
          & \cellcolor{hlblue} 4.01\% $\uparrow$ \\
          
          & HR@50  
          & 0.4482  & 0.3778 
          & 0.4224 & 0.4719  & 0.5207 
          & 0.5498 & 0.5473
          & \underline{\textcolor{baselinegray}{0.5599}} 
          & 0.4921 & 0.5286 & 0.5416 &  0.5502
          & \textbf{\textcolor{deepurple}{0.5849}}  
          & \cellcolor{hlblue} 4.47\% $\uparrow$ \\

        \midrule

    \multirow{6}{*}{\shortstack{Amazon\\Books}}
          & R@20
          & 0.0433  & 0.0539
          & 0.0597 &  0.0690 & 0.0682
          & 0.0840 & 0.0852  
          & \underline{\textcolor{baselinegray}{0.0879}}
          & 0.0713 & 0.0747 & 0.0865 & 0.0808
          & \textbf{\textcolor{deepurple}{0.0901}}  
          & \cellcolor{hlblue} 2.50\% $\uparrow$ \\
          
          & R@50   
          & 0.0677  & 0.0848
          & 0.0690 & 0.1053  & 0.1056
          & 0.1197 & 0.1248  
          & \underline{\textcolor{baselinegray}{0.1368}}
          & 0.1112 & 0.1123 & 0.1287 & 0.1194
          & \textbf{\textcolor{deepurple}{0.1378}}  
          & \cellcolor{hlblue} 0.73\% $\uparrow$ \\
          
          & ND@20 
          & 0.0340  & 0.0406
          & 0.0476 & 0.0527  & 0.0526
          & 0.0625 & 0.0640  
          & 0.0674
          & 0.0540 & 0.0562 & 0.0669 
          & \underline{\textcolor{baselinegray}{0.0732}}
          & \textbf{\textcolor{deepurple}{0.0815}}  
          & \cellcolor{hlblue} 11.34\% $\uparrow$ \\
          
          & ND@50  
          & 0.0390  & 0.0481
          & 0.0576 & 0.0587  & 0.0583
          & 0.0668 & 0.0693  
          & 0.0752
          & 0.0626 & 0.0634 & 0.0724 
          & \underline{\textcolor{baselinegray}{0.0810}}
          & \textbf{\textcolor{deepurple}{0.0971}}  
          & \cellcolor{hlblue} 19.88\% $\uparrow$ \\
          
          & HR@20
          & 0.0907  & 0.1108
          & 0.1240 & 0.1423  & 0.1411
          & 0.1676 & 0.1714  
          & \underline{\textcolor{baselinegray}{0.1804}}
          & 0.1504 & 0.1549 & 0.1782 & 0.1685
          & \textbf{\textcolor{deepurple}{0.1824}}  
          & \cellcolor{hlblue} 1.11 \% $\uparrow$ \\
          
          & HR@50  
          & 0.1379  & 0.1716
          & 0.1975 & 0.2059  & 0.2062
          & 0.2324 & 0.2413  
          & \underline{\textcolor{baselinegray}{0.2634}}
          & 0.2197 & 0.2221 & 0.2590 & 0.2426
          & \textbf{\textcolor{deepurple}{0.2638}}  
          & \cellcolor{hlblue} 0.15 \% $\uparrow$ \\

        \bottomrule
        \end{tabular}
        }
  \end{center}
\end{table*}

\subsection{Performance Comparisons}

\subsubsection{\textbf{Overall Comparisons}}

We draw the following key observations from the overall results in Table~\ref{tab:compare_perform}. 

First, \textbf{\emph{BIPCL} consistently achieves the best results} across all datasets and evaluation metrics. In particular, compared with the strongest baseline on each dataset, \emph{BIPCL} yields consistent relative improvements on NDCG@50, with gains of 4.95\%, 9.29\%, 14.79\%, 20.52\%, and 10.28\% on the five datasets, respectively. Second, \textbf{the performance gains are more pronounced on ranking metrics} (NDCG@N). This stems from effectively leveraging collective intent semantics, which provide reliable collaborative signals across the user population. By reducing noise from sparse individual interactions, collective intents enhance ranking discrimination, promoting highly relevant, consensus-driven items to the top positions. Finally, \textbf{among the baselines, \emph{DisMIR} generally delivers the strongest performance}, likely benefiting from its global item partition strategy, which disentangles latent item interests via clustering-based objectives. However, \emph{DisMIR} derives intent semantics mainly from static item co-occurrences, treating partitions as fixed item properties optimized through auxiliary objectives. This decouples item intent definition from dynamic user--item interactions, limiting the model’s ability to adapt intent based on the current sequence context. By contrast, \emph{BIPCL} directly fuses collaborative intent signals into item embeddings via bilateral intent enhancement, enabling interaction-aware intent injection that adaptively modulates embeddings according to sequence context and consistently yields superior performance. 

We further evaluate the \textbf{runtime and memory efficiency} of \emph{BIPCL} in Sec.~\ref{append_time}. \emph{BIPCL} incurs only a modest increase in training time while maintaining competitive inference efficiency. Moreover, its overall CPU and GPU memory usage remains comparable to that of existing baselines, without introducing excessive overhead.

\begin{figure*}[t]
    \centering
    \includegraphics[width=\textwidth]{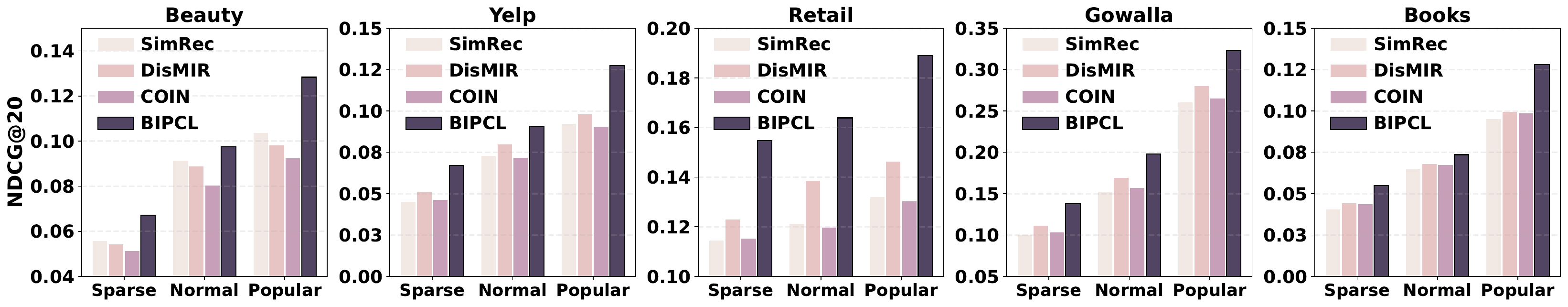}
    \caption{Performance comparison across user groups with different interaction levels (sparse, normal, and popular users).}
    \label{fig:user_group}
\end{figure*}

\subsubsection{\textbf{Comparisons w.r.t. Data Sparsity}}

To evaluate \emph{BIPCL} under varying levels of data sparsity, we partition test users into three groups according to the number of historical interactions: sparse, normal, and popular users. The results are reported in Fig.~\ref{fig:user_group}. \emph{BIPCL} consistently achieves the best performance across user groups on all datasets. While all models experience performance degradation in the sparse-user regime, \emph{BIPCL} maintains a clear and consistent advantage, indicating stronger robustness to limited interaction data. This improved robustness likely benefits from the proposed bilateral intent-enhanced modeling and multi-level contrastive regularization. By capturing shared collective patterns across items and sequences, intent-enhanced representations facilitate preference disentanglement and mitigate popularity bias, particularly benefiting sparse instances. Meanwhile, contrastive regularization alleviates representation collapse toward popular item biases under sparse supervision, enabling \emph{BIPCL} to better preserve discriminative signals when user interaction histories are limited.

\subsection{In-depth Studies of BIPCL}

\subsubsection{\textbf{Ablation Study}}
\label{abs_study}

We conduct a component-level ablation study with eight variants to evaluate key components in \emph{BIPCL}. 

\emph{(A) Intent Modeling and Fusion.} 
\textbf{\emph{w/o Pooling}} removes the average-pooled representation from the sequence encoder, i.e., $\mathbf{e}^{(u)} = \mathbf{E}^{(u)}_{T}$. 
\textbf{\emph{w/o Intent}} removes the intent enhancement module for both users and items, i.e., $\mathbf{H}^{(v)} = \mathbf{R}^{L}$ and $\mathbf{h}^{(u)} = \mathbf{e}^{(u)}$. 
\textbf{\emph{w/o Gating}} replaces gated fusion with direct aggregation, i.e., $\mathbf{H}^{(v)} = \mathbf{R}^{L} + \mathbf{Z}^{(v)}$ and $\mathbf{h}^{(u)} = \mathbf{e}^{(u)} + \mathbf{Z}^{(u)}$. 
\emph{(B) Augmentation Strategy.}
\textbf{\emph{Graph Aug}} replaces embedding-level perturbation with graph-level augmentation on the item–item co-occurrence graph, i.e., $\widetilde{\mathbf{R}}^{(1)} = (\widehat{\mathbf{A}}^{(1)})^{L} \mathbf{E}^{(v)}$ and
$\widetilde{\mathbf{R}}^{(2)} = (\widehat{\mathbf{A}}^{(2)})^{L} \mathbf{E}^{(v)}$, where $\widehat{\mathbf{A}}^{(1)}$ and $\widehat{\mathbf{A}}^{(2)}$ are independently perturbed versions of $\mathbf{A}$. 
\textbf{\emph{Seq Aug}} retains embedding-level perturbations for items, while generating two independent augmented user representations from $\mathbf{R}^{L}$ via sequence-level operations such as masking, cropping, or reordering. 
\emph{(C) Contrastive Objectives.}
\textbf{\emph{w/o CL}} removes all contrastive learning objectives by setting $\lambda=0$. 
\textbf{\emph{w/o Final-CL}} excludes contrastive alignment on final user and item representations, i.e., $\mathcal{L}_{\mathrm{CL}} =
\mathcal{L}_{\mathrm{NCE}}(\mathbf{z}^{(u,1)}, \mathbf{z}^{(u,2)})
+ \mathcal{L}_{\mathrm{NCE}}(\mathbf{z}^{(v,1)}, \mathbf{z}^{(v,2)})$. \textbf{\emph{w/o Intent-CL}} removes contrastive regularization on intent embeddings, i.e., $\mathcal{L}_{\mathrm{CL}} = 
\mathcal{L}_{\mathrm{NCE}}(\mathbf{h}^{(u,1)}, \mathbf{h}^{(u,2)})
+ \mathcal{L}_{\mathrm{NCE}}(\mathbf{h}^{(v,1)}, \mathbf{h}^{(v,2)})$.

Table~\ref{tab:abs_study_main} reports the ablation results; see Sec.~\ref{append_abs} for supplementary results. 
\emph{(A) Impact of intent modeling and fusion.} 
The performance drop in \textbf{\emph{w/o Pooling}} highlights the importance of preserving long-term user preference context beyond recency-sensitive behavioral signals. \textbf{\emph{w/o Intent}} exhibits a substantial degradation, indicating that injecting collaborative intent signals into structural embeddings is crucial for learning expressive and discriminative representations. \textbf{\emph{w/o Gating}} results in a moderate performance decline. This behavior aligns with the gradient analysis in Sec.~\ref{sec:optimization_dynamics}: although $\mathbf{J}_{\mathbf{g}}$ in Eq.~\ref{eq:grad_gating} becomes zero, the backbone still receives gradients through the residual path and can partially leverage intent information. \emph{(B) Impact of augmentation strategies.} \textbf{\emph{Graph Aug}} leads to degraded performance, as stochastic edge perturbations alter intrinsic item relations and introduce semantic drift between augmented embeddings. \textbf{\emph{Seq Aug}} also underperforms, likely because operations such as masking disrupt the temporal dependencies in user behavior sequences. Together, these results suggest that \emph{BIPCL}'s embedding perturbation yields more informative contrastive views while preserving item relations and sequential coherence. 
\emph{(C) Impact of contrastive objectives.} 
\textbf{\emph{w/o CL}} causes a sharp performance drop, confirming that contrastive supervision is critical for learning robust representations in \emph{BIPCL}. \textbf{\emph{w/o Final-CL}} results in notable regression, consistent with the analysis in Sec.~\ref{sec:uniformity}, where instance-level alignment promotes well-dispersed embeddings and alleviates over-concentration along popular directions. \textbf{\emph{w/o Intent-CL}} leads to a smaller but consistent decline. While the recommendation objective implicitly regularizes the intent space (Sec.~\ref{sec:intent_distribution}), explicit intent-level contrastive learning further improves intent discrimination and reduces redundancy among latent interests.

\begin{table}[t]
  \caption{Ablation results of \emph{BIPCL} on different datasets.}
  \label{tab:abs_study_main}
  \centering
  \setlength{\tabcolsep}{3pt}
  \resizebox{\columnwidth}{!}{
  \begin{tabular}{lcccccc}
    \toprule
    \multirow{2}{*}{Model} 
    & \multicolumn{2}{c}{Beauty} 
    & \multicolumn{2}{c}{Yelp} 
    & \multicolumn{2}{c}{Retail Rocket} \\
    \cmidrule(lr){2-3} \cmidrule(lr){4-5} \cmidrule(lr){6-7}
    & R@20 & ND@20 & R@20 & ND@20 & R@20 & ND@20 \\
    \midrule
    w/o Intent       
    & 0.1118 & 0.0771 
    & 0.1067 & 0.0847 
    & 0.2206 & 0.1326 \\
    w/o Gating       
    & 0.1222 & 0.0883 
    & 0.1174 & 0.0951 
    & 0.2559 & 0.1633 \\
    w/o Pooling      
    & 0.1186 & 0.0820 
    & 0.1142 & 0.0885
    & 0.2442 & 0.1492 \\
    Graph Aug        
    & 0.1217 & 0.0874 
    & 0.1160 & 0.0909 
    & 0.2533 & 0.1575 \\
    Seq Aug          
    & 0.1204 & 0.0852 
    & 0.1158 & 0.0895 
    & 0.2518 & 0.1523 \\
    w/o CL    
    & 0.1115 & 0.0867 
    &  0.1106 &  0.0898 
    & 0.2394 & 0.1502 \\
    w/o Final-CL     
    & 0.1198 & 0.0888 
    & 0.1133 & 0.0906 
    & 0.2486 & 0.1577 \\
    w/o Intent-CL    
    & 0.1230 & 0.0896 
    & 0.1175 & 0.0935 
    & 0.2569 & 0.1621 \\
    \midrule
    \textbf{BIPCL}    
    & \textbf{\textcolor{deepurple}{0.1277}}  
    & \textbf{\textcolor{deepurple}{0.0909}} 
    & \textbf{\textcolor{deepurple}{0.1228}} 
    & \textbf{\textcolor{deepurple}{0.0963}} 
    & \textbf{\textcolor{deepurple}{0.2580}} 
    & \textbf{\textcolor{deepurple}{0.1649}}  \\
    \bottomrule
  \end{tabular}
  }
\end{table}

\begin{figure}[t]
    \centering
    \includegraphics[width=\columnwidth]{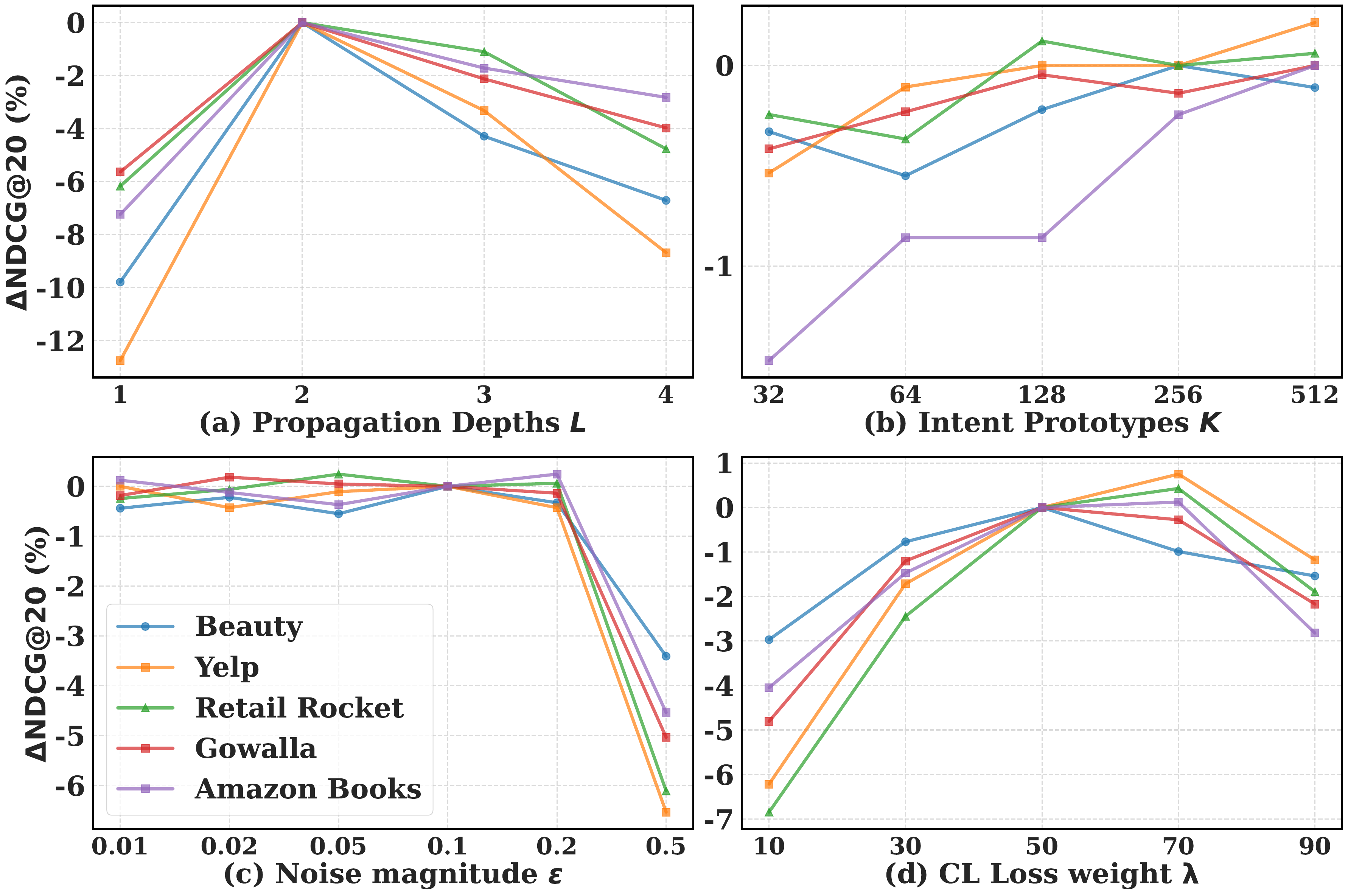}
    \caption{Hyperparameter sensitivity of \emph{BIPCL}. $\boldsymbol{\Delta} \mathrm{NDCG@20}$ denotes the relative gain over the default configuration.}
    \label{fig:params}
\end{figure}

\subsubsection{\textbf{Hyperparameter Sensitivities}}
\label{sec_params}

Fig.~\ref{fig:params} presents the performance trends of \emph{BIPCL} under varying hyperparameter settings. 
\textbf{\emph{Propagation depth $L$.}} 
As shown in Fig.~\ref{fig:params}(a), increasing $L$ from 1 to 2 yields consistent gains, indicating that \emph{BIPCL} effectively exploits multi-hop item co-occurrence to capture higher-order structural dependencies. Further increasing $L$ brings limited benefit and may even degrade performance due to over-smoothing and noise accumulation. \textbf{\emph{The number of intent prototypes $K$.}} Fig.~\ref{fig:params}(b) shows that performance improves with larger $K$, confirming the benefit of a richer intent space for modeling diverse collaborative intent patterns. However, the gains gradually diminish for large $K$, implying a capacity–redundancy trade-off where excessive prototypes lead to overlapping or weakly distinguishable intents. \textbf{\emph{Contrastive regularization strength $\lambda$.}} As shown in Fig.~\ref{fig:params}(c), moderate values of $\lambda$ consistently improves performance, suggesting that contrastive regularization promotes a better-conditioned and more uniform representation space. In contrast, overly large $\lambda$ causes performance degradation, as excessive regularization interferes with discriminative modeling and harms recommendation accuracy. 
\textbf{\emph{Perturbation magnitude $\varepsilon$.}} Fig.~\ref{fig:params}(d) indicates that \emph{BIPCL} remains robust over a broad range of perturbation magnitudes. Performance is stable for $\varepsilon \in [0.01, 0.2]$, with noticeable degradation only at large values ($\varepsilon=0.5$), where excessive noise distorts semantic information. This robustness suggests that \emph{BIPCL} is insensitive to precise perturbation tuning, supporting the stability of its contrastive view generation.

Sec.~\ref{sec:appendix_params} analyzes the sensitivity of \emph{BIPCL} to $\mathrm{TransEnc}(\cdot)$ configurations, the co-occurrence threshold $\delta$, and temperature parameters $\tau_1$ and $\tau_2$, showing strong robustness to hyperparameter choices.

\subsubsection{\textbf{Qualitative Analysis of Contrastive Learning}}
\label{sec:intent_distribution}

\begin{figure}[t]
    \centering
    \includegraphics[width=\columnwidth]{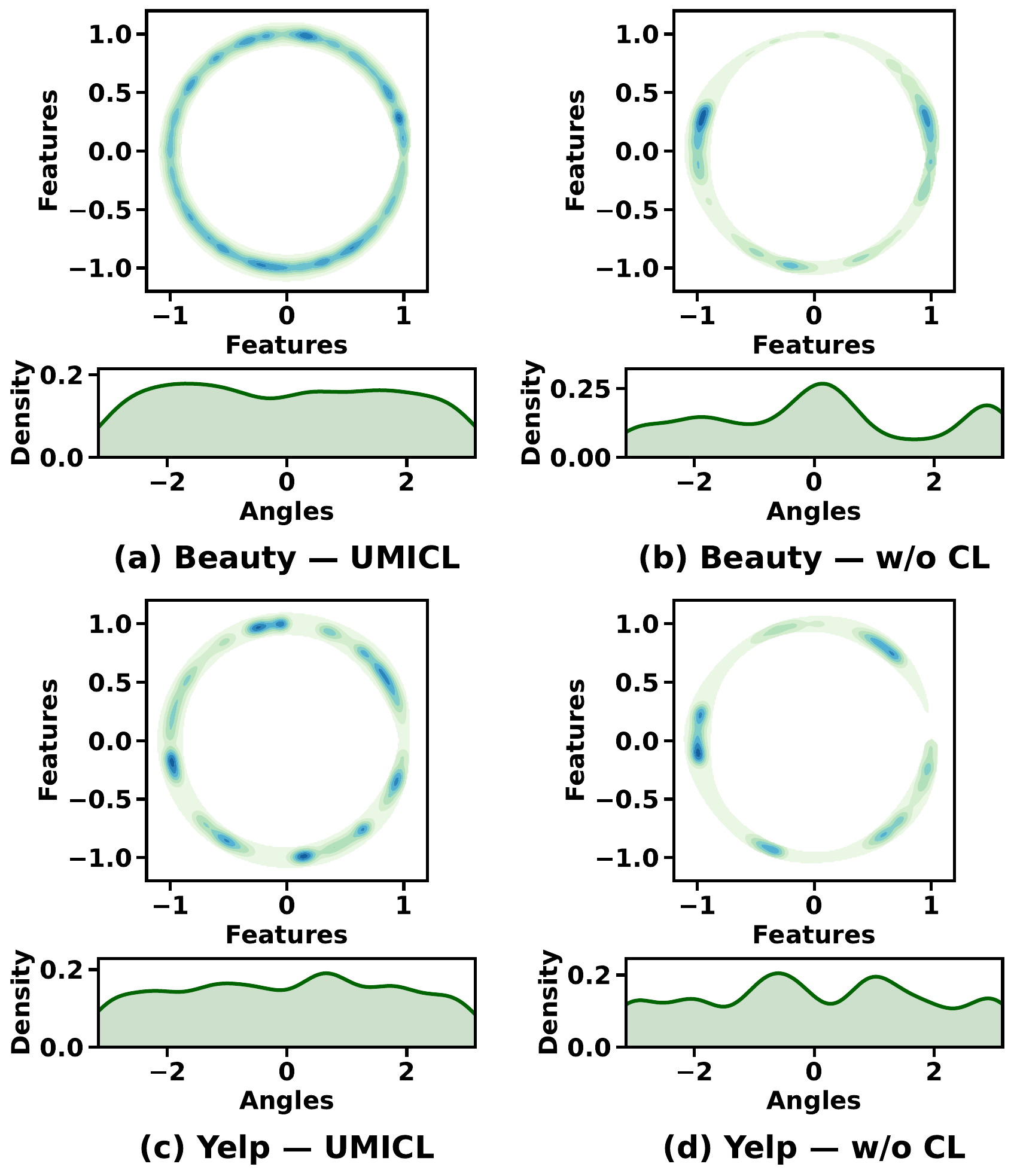}
    \caption{Density distributions of item intent embeddings.}
    \label{fig:intent_distribution}
\end{figure}

To assess the effectiveness of contrastive learning, we analyze the geometric structure of learned intent representations $\mathbf{Z}^{(v)}$ in \emph{BIPCL} and its \textbf{w/o CL} variant. Embeddings are projected into the angular space, and density distributions are estimated via kernel density estimation~\cite{draw_angles}. Implementation details are provided in the released code, and the results are shown in Fig.~\ref{fig:intent_distribution}. 
\textbf{\emph{Dispersed intent geometry in BIPCL.}} \emph{BIPCL} exhibits a well-dispersed angular layout, characterized by relatively smooth density curves without pronounced peaks. This suggests a more balanced distribution of intent semantics in the embedding space, consistent with the uniformity-promoting effect of contrastive learning discussed in Sec.~\ref{sec:uniformity}. 
\textbf{\emph{Intent concentration in w/o CL.}}
In contrast, \textbf{\emph{w/o CL}} shows localized concentration in specific angular regions, reflected by moderate peaks in the density curves. Such geometric concentration suggests increased redundancy among intent embeddings. This phenomenon may partially account for the observed performance drop of \textbf{\emph{w/o CL}} in Table~\ref{tab:abs_study_main}, as overlapping intents limit fine-grained preference discrimination. 
Overall, the comparison indicates that contrastive learning regularizes intent geometry by promoting angular dispersion and reducing redundancy, correlating with \emph{BIPCL}'s improved performance~\cite{simgcl}.

\begin{figure}[t]
    \centering
    \includegraphics[width=\columnwidth]{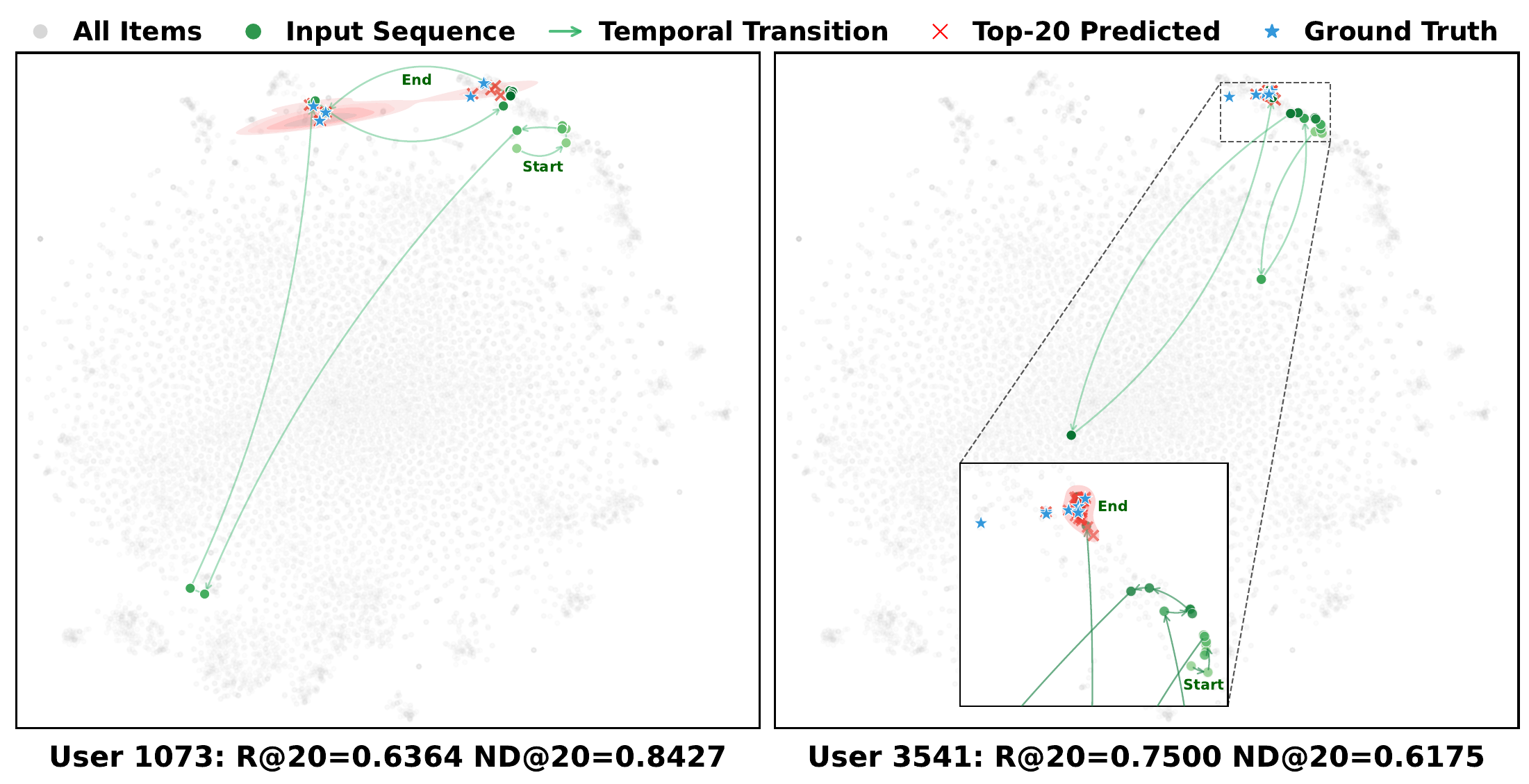}
    \caption{Visualization of user interaction and recommendation embeddings. The complete code and data are available.}
    \label{fig:case_study}
\end{figure}

\subsubsection{\textbf{Case Study}}
\label{sec:case_study}

We conduct a case study on the Beauty dataset to examine how \emph{BIPCL} responds to evolving user preferences and whether its recommendations remain semantically consistent with future interactions. Final item representations $\mathbf{H}^{(v)}$ are projected into two-dimensional space using t-SNE. We select two representative users and visualize their historical interactions, ground-truth (GT) future items, and \emph{BIPCL}'s top-20 recommendations in the shared embedding space (Fig.~\ref{fig:case_study}). 
For User 1073, whose preferences shift abruptly across item regions, the recommendations form two distinct clusters aligned with both recent and earlier interactions. Their close overlap with GT items suggests that \emph{BIPCL} remains sensitive to multiple preference signals rather than overemphasizing recency. 
In contrast, User 3541 exhibits smoother preference evolution, with recommendations concentrated along the dominant interaction trajectory and tightly aligned with GT items, consistent with a coherent progression of user intents. 
This visual alignment indicates that \emph{BIPCL}'s recommendations are informed by both short- and long-term behavioral signals. By contrast, \textbf{\emph{w/o Pooling}} relies solely on the final hidden state as the sequence summary, which biases representations toward recent behavior. This design weakens the contribution of earlier interactions, potentially explaining its pronounced performance degradation.

\section{CONCLUSION}

We propose \emph{BIPCL}, an end-to-end contrastive learning framework for multi-intent SR. \emph{BIPCL} explicitly injects multi-intent signals into both item and temporal sequence representations via a bilateral intent-enhancement mechanism. \emph{BIPCL} further constructs effective contrastive views by perturbing structural item embeddings, and performs multi-level alignment across interaction- and intent-level representations. Extensive experiments show that \emph{BIPCL} consistently outperforms state-of-the-art baselines. Ablation studies verify the effectiveness of the core mechanisms in \emph{BIPCL}, including intent enhancement and multi-level contrastive alignment.


\bibliographystyle{ACM-Reference-Format}
\bibliography{sample-base}


\begin{thebibliography}{58}


\ifx \showCODEN    \undefined \def \showCODEN     #1{\unskip}     \fi
\ifx \showISBNx    \undefined \def \showISBNx     #1{\unskip}     \fi
\ifx \showISBNxiii \undefined \def \showISBNxiii  #1{\unskip}     \fi
\ifx \showISSN     \undefined \def \showISSN      #1{\unskip}     \fi
\ifx \showLCCN     \undefined \def \showLCCN      #1{\unskip}     \fi
\ifx \shownote     \undefined \def \shownote      #1{#1}          \fi
\ifx \showarticletitle \undefined \def \showarticletitle #1{#1}   \fi
\ifx \showURL      \undefined \def \showURL       {\relax}        \fi
\providecommand\bibfield[2]{#2}
\providecommand\bibinfo[2]{#2}
\providecommand\natexlab[1]{#1}
\providecommand\showeprint[2][]{arXiv:#2}

\bibitem[Bai et~al\mbox{.}(2018)]%
        {aux_signs_1}
\bibfield{author}{\bibinfo{person}{Ting Bai}, \bibinfo{person}{Jian-Yun Nie}, \bibinfo{person}{Wayne~Xin Zhao}, \bibinfo{person}{Yutao Zhu}, \bibinfo{person}{Pan Du}, {and} \bibinfo{person}{Ji-Rong Wen}.} \bibinfo{year}{2018}\natexlab{}.
\newblock \showarticletitle{An Attribute-aware Neural Attentive Model for Next Basket Recommendation}. In \bibinfo{booktitle}{\emph{The 41st International ACM SIGIR Conference on Research \& Development in Information Retrieval}} \emph{(\bibinfo{series}{SIGIR '18})}. \bibinfo{publisher}{Association for Computing Machinery}, \bibinfo{address}{New York, NY, USA}, \bibinfo{pages}{1201--1204}.
\newblock
\href{https://doi.org/10.1145/3209978.3210129}{doi:\nolinkurl{10.1145/3209978.3210129}}


\bibitem[Cen et~al\mbox{.}(2020)]%
        {comirec}
\bibfield{author}{\bibinfo{person}{Yukuo Cen}, \bibinfo{person}{Jianwei Zhang}, \bibinfo{person}{Xu Zou}, \bibinfo{person}{Chang Zhou}, \bibinfo{person}{Hongxia Yang}, {and} \bibinfo{person}{Jie Tang}.} \bibinfo{year}{2020}\natexlab{}.
\newblock \showarticletitle{Controllable Multi-Interest Framework for Recommendation}. In \bibinfo{booktitle}{\emph{Proceedings of the 26th ACM SIGKDD International Conference on Knowledge Discovery \& Data Mining}} \emph{(\bibinfo{series}{KDD '20})}. \bibinfo{publisher}{Association for Computing Machinery}, \bibinfo{address}{New York, NY, USA}, \bibinfo{pages}{2942--2951}.
\newblock
\href{https://doi.org/10.1145/3394486.3403344}{doi:\nolinkurl{10.1145/3394486.3403344}}


\bibitem[Chai et~al\mbox{.}(2022)]%
        {umi}
\bibfield{author}{\bibinfo{person}{Zheng Chai}, \bibinfo{person}{Zhihong Chen}, \bibinfo{person}{Chenliang Li}, \bibinfo{person}{Rong Xiao}, \bibinfo{person}{Houyi Li}, \bibinfo{person}{Jiawei Wu}, \bibinfo{person}{Jingxu Chen}, {and} \bibinfo{person}{Haihong Tang}.} \bibinfo{year}{2022}\natexlab{}.
\newblock \showarticletitle{User-Aware Multi-Interest Learning for Candidate Matching in Recommenders}. In \bibinfo{booktitle}{\emph{Proceedings of the 45th International ACM SIGIR Conference on Research and Development in Information Retrieval}} \emph{(\bibinfo{series}{SIGIR '22})}. \bibinfo{publisher}{Association for Computing Machinery}, \bibinfo{address}{New York, NY, USA}, \bibinfo{pages}{1326--1335}.
\newblock
\href{https://doi.org/10.1145/3477495.3532073}{doi:\nolinkurl{10.1145/3477495.3532073}}


\bibitem[Chen et~al\mbox{.}(2024)]%
        {intent_works_1}
\bibfield{author}{\bibinfo{person}{Gaode Chen}, \bibinfo{person}{Yuezihan Jiang}, \bibinfo{person}{Rui Huang}, \bibinfo{person}{Kuo Cai}, \bibinfo{person}{Yunze Luo}, \bibinfo{person}{Ruina Sun}, \bibinfo{person}{Qi Zhang}, \bibinfo{person}{Han Li}, {and} \bibinfo{person}{Kun Gai}.} \bibinfo{year}{2024}\natexlab{}.
\newblock \showarticletitle{Missing Interest Modeling with Lifelong User Behavior Data for Retrieval Recommendation}. In \bibinfo{booktitle}{\emph{Proceedings of the 33rd ACM International Conference on Information and Knowledge Management}} \emph{(\bibinfo{series}{CIKM '24})}. \bibinfo{publisher}{Association for Computing Machinery}, \bibinfo{address}{New York, NY, USA}, \bibinfo{pages}{4390--4396}.
\newblock
\href{https://doi.org/10.1145/3627673.3680019}{doi:\nolinkurl{10.1145/3627673.3680019}}


\bibitem[Chen et~al\mbox{.}(2021b)]%
        {pimirec}
\bibfield{author}{\bibinfo{person}{Gaode Chen}, \bibinfo{person}{Xinghua Zhang}, \bibinfo{person}{Yanyan Zhao}, \bibinfo{person}{Cong Xue}, {and} \bibinfo{person}{ji Xiang}.} \bibinfo{year}{2021}\natexlab{b}.
\newblock \showarticletitle{Exploring Periodicity and Interactivity in Multi-Interest Framework for Sequential Recommendation}. In \bibinfo{booktitle}{\emph{Proceedings of the Thirtieth International Joint Conference on Artificial Intelligence}} \emph{(\bibinfo{series}{IJCAI '21})}. \bibinfo{pages}{1426--1433}.
\newblock


\bibitem[Chen et~al\mbox{.}(2021a)]%
        {aux_rec_2}
\bibfield{author}{\bibinfo{person}{Hong Chen}, \bibinfo{person}{Yudong Chen}, \bibinfo{person}{Xin Wang}, \bibinfo{person}{Ruobing Xie}, \bibinfo{person}{Rui Wang}, \bibinfo{person}{Feng Xia}, {and} \bibinfo{person}{Wenwu Zhu}.} \bibinfo{year}{2021}\natexlab{a}.
\newblock \showarticletitle{Curriculum disentangled recommendation with noisy multi-feedback}. In \bibinfo{booktitle}{\emph{Proceedings of the 35th International Conference on Neural Information Processing Systems}} \emph{(\bibinfo{series}{NIPS '21})}. \bibinfo{publisher}{Curran Associates Inc.}
\newblock


\bibitem[Chen et~al\mbox{.}(2020)]%
        {infonce}
\bibfield{author}{\bibinfo{person}{Ting Chen}, \bibinfo{person}{Simon Kornblith}, \bibinfo{person}{Mohammad Norouzi}, {and} \bibinfo{person}{Geoffrey Hinton}.} \bibinfo{year}{2020}\natexlab{}.
\newblock \showarticletitle{A Simple Framework for Contrastive Learning of Visual Representations}. In \bibinfo{booktitle}{\emph{Proceedings of the 37th International Conference on Machine Learning}} \emph{(\bibinfo{series}{ICML '20})}. \bibinfo{publisher}{JMLR.org}.
\newblock


\bibitem[Chen et~al\mbox{.}(2018)]%
        {memory_1}
\bibfield{author}{\bibinfo{person}{Xu Chen}, \bibinfo{person}{Hongteng Xu}, \bibinfo{person}{Yongfeng Zhang}, \bibinfo{person}{Jiaxi Tang}, \bibinfo{person}{Yixin Cao}, \bibinfo{person}{Zheng Qin}, {and} \bibinfo{person}{Hongyuan Zha}.} \bibinfo{year}{2018}\natexlab{}.
\newblock \showarticletitle{Sequential Recommendation with User Memory Networks}. In \bibinfo{booktitle}{\emph{Proceedings of the Eleventh ACM International Conference on Web Search and Data Mining}} \emph{(\bibinfo{series}{WSDM '18})}. \bibinfo{publisher}{Association for Computing Machinery}, \bibinfo{address}{New York, NY, USA}, \bibinfo{pages}{108--116}.
\newblock
\href{https://doi.org/10.1145/3159652.3159668}{doi:\nolinkurl{10.1145/3159652.3159668}}


\bibitem[Chen et~al\mbox{.}(2022)]%
        {iclrec}
\bibfield{author}{\bibinfo{person}{Yongjun Chen}, \bibinfo{person}{Zhiwei Liu}, \bibinfo{person}{Jia Li}, \bibinfo{person}{Julian McAuley}, {and} \bibinfo{person}{Caiming Xiong}.} \bibinfo{year}{2022}\natexlab{}.
\newblock \showarticletitle{Intent Contrastive Learning for Sequential Recommendation}. In \bibinfo{booktitle}{\emph{Proceedings of the ACM Web Conference 2022}} \emph{(\bibinfo{series}{WWW '22})}. \bibinfo{publisher}{Association for Computing Machinery}, \bibinfo{address}{New York, NY, USA}, \bibinfo{pages}{2172--2182}.
\newblock
\href{https://doi.org/10.1145/3485447.3512090}{doi:\nolinkurl{10.1145/3485447.3512090}}


\bibitem[Cheng et~al\mbox{.}(2025)]%
        {coin}
\bibfield{author}{\bibinfo{person}{Yu-Ting Cheng}, \bibinfo{person}{Yu-Yen Ho}, {and} \bibinfo{person}{Jyun-Yu Jiang}.} \bibinfo{year}{2025}\natexlab{}.
\newblock \showarticletitle{Collaborative Interest Modeling in Recommender Systems}. In \bibinfo{booktitle}{\emph{Proceedings of the Nineteenth ACM Conference on Recommender Systems}} \emph{(\bibinfo{series}{RecSys '25})}. \bibinfo{publisher}{Association for Computing Machinery}, \bibinfo{address}{New York, NY, USA}, \bibinfo{pages}{533--538}.
\newblock
\href{https://doi.org/10.1145/3705328.3748023}{doi:\nolinkurl{10.1145/3705328.3748023}}


\bibitem[Dang et~al\mbox{.}(2024)]%
        {TiCoSeRec}
\bibfield{author}{\bibinfo{person}{Yizhou Dang}, \bibinfo{person}{Enneng Yang}, \bibinfo{person}{Guibing Guo}, \bibinfo{person}{Linying Jiang}, \bibinfo{person}{Xingwei Wang}, \bibinfo{person}{Xiaoxiao Xu}, \bibinfo{person}{Qinghui Sun}, {and} \bibinfo{person}{Hong Liu}.} \bibinfo{year}{2024}\natexlab{}.
\newblock \showarticletitle{TiCoSeRec: Augmenting Data to Uniform Sequences by Time Intervals for Effective Recommendation}.
\newblock \bibinfo{journal}{\emph{IEEE Transactions on Knowledge and Data Engineering}} \bibinfo{volume}{36}, \bibinfo{number}{6} (\bibinfo{year}{2024}), \bibinfo{pages}{2686--2700}.
\newblock
\href{https://doi.org/10.1109/TKDE.2023.3324312}{doi:\nolinkurl{10.1109/TKDE.2023.3324312}}


\bibitem[Donkers et~al\mbox{.}(2017)]%
        {rnn}
\bibfield{author}{\bibinfo{person}{Tim Donkers}, \bibinfo{person}{Benedikt Loepp}, {and} \bibinfo{person}{J\'{u}rgen Ziegler}.} \bibinfo{year}{2017}\natexlab{}.
\newblock \showarticletitle{Sequential User-based Recurrent Neural Network Recommendations}. In \bibinfo{booktitle}{\emph{Proceedings of the Eleventh ACM Conference on Recommender Systems}} \emph{(\bibinfo{series}{RecSys '17})}. \bibinfo{publisher}{Association for Computing Machinery}, \bibinfo{address}{New York, NY, USA}, \bibinfo{pages}{152--160}.
\newblock
\href{https://doi.org/10.1145/3109859.3109877}{doi:\nolinkurl{10.1145/3109859.3109877}}


\bibitem[Du et~al\mbox{.}(2024)]%
        {dismir}
\bibfield{author}{\bibinfo{person}{Yingpeng Du}, \bibinfo{person}{Ziyan Wang}, \bibinfo{person}{Zhu Sun}, \bibinfo{person}{Yining Ma}, \bibinfo{person}{Hongzhi Liu}, {and} \bibinfo{person}{Jie Zhang}.} \bibinfo{year}{2024}\natexlab{}.
\newblock \showarticletitle{Disentangled Multi-interest Representation Learning for Sequential Recommendation}. In \bibinfo{booktitle}{\emph{Proceedings of the 30th ACM SIGKDD Conference on Knowledge Discovery and Data Mining}} \emph{(\bibinfo{series}{KDD '24})}. \bibinfo{publisher}{Association for Computing Machinery}, \bibinfo{address}{New York, NY, USA}, \bibinfo{pages}{677--688}.
\newblock
\href{https://doi.org/10.1145/3637528.3671800}{doi:\nolinkurl{10.1145/3637528.3671800}}


\bibitem[Fang et~al\mbox{.}(2020)]%
        {srec_survey}
\bibfield{author}{\bibinfo{person}{Hui Fang}, \bibinfo{person}{Danning Zhang}, \bibinfo{person}{Yiheng Shu}, {and} \bibinfo{person}{Guibing Guo}.} \bibinfo{year}{2020}\natexlab{}.
\newblock \showarticletitle{Deep Learning for Sequential Recommendation: Algorithms, Influential Factors, and Evaluations}.
\newblock \bibinfo{journal}{\emph{ACM Trans. Inf. Syst.}} \bibinfo{volume}{39}, \bibinfo{number}{1} (\bibinfo{year}{2020}), \bibinfo{pages}{1--42}.
\newblock
\href{https://doi.org/10.1145/3426723}{doi:\nolinkurl{10.1145/3426723}}


\bibitem[Goodfellow et~al\mbox{.}(2015)]%
        {fgsm}
\bibfield{author}{\bibinfo{person}{Ian~J. Goodfellow}, \bibinfo{person}{Jonathon Shlens}, {and} \bibinfo{person}{Christian Szegedy}.} \bibinfo{year}{2015}\natexlab{}.
\newblock \showarticletitle{Explaining and Harnessing Adversarial Examples}. In \bibinfo{booktitle}{\emph{International Conference on Learning Representations}} \emph{(\bibinfo{series}{ICLR '15})}.
\newblock


\bibitem[He and McAuley(2016)]%
        {mc_1}
\bibfield{author}{\bibinfo{person}{Ruining He} {and} \bibinfo{person}{Julian McAuley}.} \bibinfo{year}{2016}\natexlab{}.
\newblock \showarticletitle{Fusing Similarity Models with Markov Chains for Sparse Sequential Recommendation}. In \bibinfo{booktitle}{\emph{2016 IEEE 16th International Conference on Data Mining (ICDM)}} \emph{(\bibinfo{series}{ICDM '16})}. \bibinfo{pages}{191--200}.
\newblock
\href{https://doi.org/10.1109/ICDM.2016.0030}{doi:\nolinkurl{10.1109/ICDM.2016.0030}}


\bibitem[He et~al\mbox{.}(2025)]%
        {intent_works_2}
\bibfield{author}{\bibinfo{person}{Zhiyu He}, \bibinfo{person}{Zhixin Ling}, \bibinfo{person}{Jiayu Li}, \bibinfo{person}{Zhiqiang Guo}, \bibinfo{person}{Weizhi Ma}, \bibinfo{person}{Xinchen Luo}, \bibinfo{person}{Min Zhang}, {and} \bibinfo{person}{Guorui Zhou}.} \bibinfo{year}{2025}\natexlab{}.
\newblock \showarticletitle{Short Video Segment-level User Dynamic Interests Modeling in Personalized Recommendation}. In \bibinfo{booktitle}{\emph{Proceedings of the 48th International ACM SIGIR Conference on Research and Development in Information Retrieval}} \emph{(\bibinfo{series}{SIGIR '25})}. \bibinfo{publisher}{Association for Computing Machinery}, \bibinfo{address}{New York, NY, USA}, \bibinfo{pages}{1880--1890}.
\newblock
\href{https://doi.org/10.1145/3726302.3730083}{doi:\nolinkurl{10.1145/3726302.3730083}}


\bibitem[Huang et~al\mbox{.}(2018)]%
        {aux_signs_2}
\bibfield{author}{\bibinfo{person}{Jin Huang}, \bibinfo{person}{Wayne~Xin Zhao}, \bibinfo{person}{Hongjian Dou}, \bibinfo{person}{Ji-Rong Wen}, {and} \bibinfo{person}{Edward~Y. Chang}.} \bibinfo{year}{2018}\natexlab{}.
\newblock \showarticletitle{Improving Sequential Recommendation with Knowledge-Enhanced Memory Networks}. In \bibinfo{booktitle}{\emph{The 41st International ACM SIGIR Conference on Research \& Development in Information Retrieval}} \emph{(\bibinfo{series}{SIGIR '18})}. \bibinfo{publisher}{Association for Computing Machinery}, \bibinfo{address}{New York, NY, USA}, \bibinfo{pages}{505--514}.
\newblock
\href{https://doi.org/10.1145/3209978.3210017}{doi:\nolinkurl{10.1145/3209978.3210017}}


\bibitem[Jean et~al\mbox{.}(2015)]%
        {sampled_softmax}
\bibfield{author}{\bibinfo{person}{S{\'e}bastien Jean}, \bibinfo{person}{Kyunghyun Cho}, \bibinfo{person}{Roland Memisevic}, {and} \bibinfo{person}{Yoshua Bengio}.} \bibinfo{year}{2015}\natexlab{}.
\newblock \showarticletitle{On Using Very Large Target Vocabulary for Neural Machine Translation}. In \bibinfo{booktitle}{\emph{Proceedings of the 53rd Annual Meeting of the Association for Computational Linguistics and the 7th International Joint Conference on Natural Language Processing (Volume 1: Long Papers)}} \emph{(\bibinfo{series}{ACL '15})}. \bibinfo{publisher}{Association for Computational Linguistics}, \bibinfo{pages}{1--10}.
\newblock
\href{https://doi.org/10.3115/v1/P15-1001}{doi:\nolinkurl{10.3115/v1/P15-1001}}


\bibitem[Jing et~al\mbox{.}(2023)]%
        {cl_survey}
\bibfield{author}{\bibinfo{person}{Mengyuan Jing}, \bibinfo{person}{Yanmin Zhu}, \bibinfo{person}{Tianzi Zang}, {and} \bibinfo{person}{Ke Wang}.} \bibinfo{year}{2023}\natexlab{}.
\newblock \showarticletitle{Contrastive Self-supervised Learning in Recommender Systems: A Survey}.
\newblock \bibinfo{journal}{\emph{ACM Trans. Inf. Syst.}} \bibinfo{volume}{42}, \bibinfo{number}{2} (\bibinfo{year}{2023}), \bibinfo{pages}{1--39}.
\newblock
\href{https://doi.org/10.1145/3627158}{doi:\nolinkurl{10.1145/3627158}}


\bibitem[Ju et~al\mbox{.}(2025)]%
        {cross_domain_2}
\bibfield{author}{\bibinfo{person}{Clark~Mingxuan Ju}, \bibinfo{person}{Leonardo Neves}, \bibinfo{person}{Bhuvesh Kumar}, \bibinfo{person}{Liam Collins}, \bibinfo{person}{Tong Zhao}, \bibinfo{person}{Yuwei Qiu}, \bibinfo{person}{Qing Dou}, \bibinfo{person}{Sohail Nizam}, \bibinfo{person}{Sen Yang}, {and} \bibinfo{person}{Neil Shah}.} \bibinfo{year}{2025}\natexlab{}.
\newblock \showarticletitle{Revisiting Self-attention for Cross-domain Sequential Recommendation}. In \bibinfo{booktitle}{\emph{Proceedings of the 31st ACM SIGKDD Conference on Knowledge Discovery and Data Mining V.2}} \emph{(\bibinfo{series}{KDD '25})}. \bibinfo{publisher}{Association for Computing Machinery}, \bibinfo{address}{New York, NY, USA}, \bibinfo{pages}{1094--1105}.
\newblock
\href{https://doi.org/10.1145/3711896.3737108}{doi:\nolinkurl{10.1145/3711896.3737108}}


\bibitem[Kang and McAuley(2018)]%
        {sasrec}
\bibfield{author}{\bibinfo{person}{Wang-Cheng Kang} {and} \bibinfo{person}{Julian McAuley}.} \bibinfo{year}{2018}\natexlab{}.
\newblock \showarticletitle{Self-Attentive Sequential Recommendation}. In \bibinfo{booktitle}{\emph{2018 IEEE International Conference on Data Mining (ICDM)}} \emph{(\bibinfo{series}{ICDM '18})}. \bibinfo{pages}{197--206}.
\newblock
\href{https://doi.org/10.1109/ICDM.2018.00035}{doi:\nolinkurl{10.1109/ICDM.2018.00035}}


\bibitem[level Deeper Self-Attention Network~for Sequential~Recommendation(2019)]%
        {fdsa}
\bibfield{author}{\bibinfo{person}{Feature level Deeper Self-Attention Network~for Sequential~Recommendation}.} \bibinfo{year}{2019}\natexlab{}.
\newblock \showarticletitle{Zhang, Tingting and Zhao, Pengpeng and Liu, Yanchi and Sheng, Victor S. and Xu, Jiajie and Wang, Deqing and Liu, Guanfeng and Zhou, Xiaofang}. In \bibinfo{booktitle}{\emph{Proceedings of the Twenty-Eighth International Joint Conference on Artificial Intelligence, {IJCAI-19}}}. \bibinfo{publisher}{International Joint Conferences on Artificial Intelligence Organization}, \bibinfo{pages}{4320--4326}.
\newblock
\href{https://doi.org/10.24963/ijcai.2019/600}{doi:\nolinkurl{10.24963/ijcai.2019/600}}


\bibitem[Li et~al\mbox{.}(2019)]%
        {mind}
\bibfield{author}{\bibinfo{person}{Chao Li}, \bibinfo{person}{Zhiyuan Liu}, \bibinfo{person}{Mengmeng Wu}, \bibinfo{person}{Yuchi Xu}, \bibinfo{person}{Huan Zhao}, \bibinfo{person}{Pipei Huang}, \bibinfo{person}{Guoliang Kang}, \bibinfo{person}{Qiwei Chen}, \bibinfo{person}{Wei Li}, {and} \bibinfo{person}{Dik~Lun Lee}.} \bibinfo{year}{2019}\natexlab{}.
\newblock \showarticletitle{Multi-Interest Network with Dynamic Routing for Recommendation at Tmall}. In \bibinfo{booktitle}{\emph{Proceedings of the 28th ACM International Conference on Information and Knowledge Management}} \emph{(\bibinfo{series}{CIKM '19})}. \bibinfo{publisher}{Association for Computing Machinery}, \bibinfo{address}{New York, NY, USA}, \bibinfo{pages}{2615--2623}.
\newblock
\href{https://doi.org/10.1145/3357384.3357814}{doi:\nolinkurl{10.1145/3357384.3357814}}


\bibitem[Li et~al\mbox{.}(2022)]%
        {aux_rec_1}
\bibfield{author}{\bibinfo{person}{Jian Li}, \bibinfo{person}{Jieming Zhu}, \bibinfo{person}{Qiwei Bi}, \bibinfo{person}{Guohao Cai}, \bibinfo{person}{Lifeng Shang}, \bibinfo{person}{Zhenhua Dong}, \bibinfo{person}{Xin Jiang}, {and} \bibinfo{person}{Qun Liu}.} \bibinfo{year}{2022}\natexlab{}.
\newblock \showarticletitle{MINER: Multi-Interest Matching Network for News Recommendation}. In \bibinfo{booktitle}{\emph{Findings of the Association for Computational Linguistics: ACL 2022}}. \bibinfo{publisher}{Association for Computational Linguistics}, \bibinfo{pages}{343--352}.
\newblock
\urldef\tempurl%
\url{https://aclanthology.org/2022.findings-acl.29/}
\showURL{%
\tempurl}


\bibitem[Li et~al\mbox{.}(2024)]%
        {frec}
\bibfield{author}{\bibinfo{person}{Nian Li}, \bibinfo{person}{Xin Ban}, \bibinfo{person}{Cheng Ling}, \bibinfo{person}{Chen Gao}, \bibinfo{person}{Lantao Hu}, \bibinfo{person}{Peng Jiang}, \bibinfo{person}{Kun Gai}, \bibinfo{person}{Yong Li}, {and} \bibinfo{person}{Qingmin Liao}.} \bibinfo{year}{2024}\natexlab{}.
\newblock \showarticletitle{Modeling User Fatigue for Sequential Recommendation}. In \bibinfo{booktitle}{\emph{Proceedings of the 47th International ACM SIGIR Conference on Research and Development in Information Retrieval}} \emph{(\bibinfo{series}{SIGIR '24})}. \bibinfo{publisher}{Association for Computing Machinery}, \bibinfo{address}{New York, NY, USA}, \bibinfo{pages}{996--1005}.
\newblock
\href{https://doi.org/10.1145/3626772.3657802}{doi:\nolinkurl{10.1145/3626772.3657802}}


\bibitem[Lin et~al\mbox{.}(2026)]%
        {srec_survey_2}
\bibfield{author}{\bibinfo{person}{Sitao Lin}, \bibinfo{person}{Shuai Tang}, \bibinfo{person}{Xiaofeng Zhang}, \bibinfo{person}{Jianghong Ma}, {and} \bibinfo{person}{Ziao Wang}.} \bibinfo{year}{2026}\natexlab{}.
\newblock \showarticletitle{CoDeR+: Interest-aware Counterfactual Reasoning for Sequential Recommendation}.
\newblock \bibinfo{journal}{\emph{ACM Trans. Inf. Syst.}} \bibinfo{volume}{44}, \bibinfo{number}{2} (\bibinfo{year}{2026}), \bibinfo{pages}{1--39}.
\newblock
\href{https://doi.org/10.1145/3778863}{doi:\nolinkurl{10.1145/3778863}}


\bibitem[Lin et~al\mbox{.}(2025)]%
        {intent_drift}
\bibfield{author}{\bibinfo{person}{Xiaolin Lin}, \bibinfo{person}{Weike Pan}, {and} \bibinfo{person}{Zhong Ming}.} \bibinfo{year}{2025}\natexlab{}.
\newblock \showarticletitle{Towards Interest Drift-driven User Representation Learning in Sequential Recommendation}. In \bibinfo{booktitle}{\emph{Proceedings of the 48th International ACM SIGIR Conference on Research and Development in Information Retrieval}} \emph{(\bibinfo{series}{SIGIR '25})}. \bibinfo{publisher}{Association for Computing Machinery}, \bibinfo{address}{New York, NY, USA}, \bibinfo{pages}{1541--1551}.
\newblock
\href{https://doi.org/10.1145/3726302.3730099}{doi:\nolinkurl{10.1145/3726302.3730099}}


\bibitem[Liu et~al\mbox{.}(2022)]%
        {aux_rec_4}
\bibfield{author}{\bibinfo{person}{Danyang Liu}, \bibinfo{person}{Yuji Yang}, \bibinfo{person}{Mengdi Zhang}, \bibinfo{person}{Wei Wu}, \bibinfo{person}{Xing Xie}, {and} \bibinfo{person}{Guangzhong Sun}.} \bibinfo{year}{2022}\natexlab{}.
\newblock \showarticletitle{Knowledge Enhanced Multi-Interest Network for the Generation of Recommendation Candidates}. In \bibinfo{booktitle}{\emph{Proceedings of the 31st ACM International Conference on Information \& Knowledge Management}} \emph{(\bibinfo{series}{CIKM '22})}. \bibinfo{publisher}{Association for Computing Machinery}, \bibinfo{address}{New York, NY, USA}, \bibinfo{pages}{3322--3331}.
\newblock
\href{https://doi.org/10.1145/3511808.3557114}{doi:\nolinkurl{10.1145/3511808.3557114}}


\bibitem[Liu et~al\mbox{.}(2024a)]%
        {simrec}
\bibfield{author}{\bibinfo{person}{Yaokun Liu}, \bibinfo{person}{Xiaowang Zhang}, \bibinfo{person}{Minghui Zou}, {and} \bibinfo{person}{Zhiyong Feng}.} \bibinfo{year}{2024}\natexlab{a}.
\newblock \showarticletitle{Attribute Simulation for Item Embedding Enhancement in Multi-interest Recommendation}. In \bibinfo{booktitle}{\emph{Proceedings of the 17th ACM International Conference on Web Search and Data Mining}} \emph{(\bibinfo{series}{WSDM '24})}. \bibinfo{publisher}{Association for Computing Machinery}, \bibinfo{address}{New York, NY, USA}, \bibinfo{pages}{482--491}.
\newblock
\href{https://doi.org/10.1145/3616855.3635841}{doi:\nolinkurl{10.1145/3616855.3635841}}


\bibitem[Liu et~al\mbox{.}(2024b)]%
        {elcrec}
\bibfield{author}{\bibinfo{person}{Yue Liu}, \bibinfo{person}{Shihao Zhu}, \bibinfo{person}{Jun Xia}, \bibinfo{person}{Yingwei Ma}, \bibinfo{person}{Jian Ma}, \bibinfo{person}{Xinwang Liu}, \bibinfo{person}{Shengju Yu}, \bibinfo{person}{Kejun Zhang}, {and} \bibinfo{person}{Wenliang Zhong}.} \bibinfo{year}{2024}\natexlab{b}.
\newblock \showarticletitle{End-to-end Learnable Clustering for Intent Learning in Recommendation}. In \bibinfo{booktitle}{\emph{Advances in Neural Information Processing Systems}} \emph{(\bibinfo{series}{NIPS '24})}. \bibinfo{publisher}{Curran Associates, Inc.}, \bibinfo{pages}{5913--5949}.
\newblock
\href{https://doi.org/10.52202/079017-0192}{doi:\nolinkurl{10.52202/079017-0192}}


\bibitem[Liu et~al\mbox{.}(2021)]%
        {coserec}
\bibfield{author}{\bibinfo{person}{Zhiwei Liu}, \bibinfo{person}{Yongjun Chen}, \bibinfo{person}{Jia Li}, \bibinfo{person}{Philip~S. Yu}, \bibinfo{person}{Julian McAuley}, {and} \bibinfo{person}{Caiming Xiong}.} \bibinfo{year}{2021}\natexlab{}.
\newblock \bibinfo{booktitle}{\emph{Contrastive Self-supervised Sequential Recommendation with Robust Augmentation}}.
\newblock
\showeprint[arxiv]{2108.06479}
\href{https://doi.org/10.48550/arXiv.2108.06479}{doi:\nolinkurl{10.48550/arXiv.2108.06479}}


\bibitem[Ma et~al\mbox{.}(2019)]%
        {intent_attn_1}
\bibfield{author}{\bibinfo{person}{Jianxin Ma}, \bibinfo{person}{Chang Zhou}, \bibinfo{person}{Peng Cui}, \bibinfo{person}{Hongxia Yang}, {and} \bibinfo{person}{Wenwu Zhu}.} \bibinfo{year}{2019}\natexlab{}.
\newblock \showarticletitle{Learning Disentangled Representations for Recommendation}. In \bibinfo{booktitle}{\emph{Advances in Neural Information Processing Systems}} \emph{(\bibinfo{series}{NIPS '19})}. \bibinfo{publisher}{Curran Associates, Inc.}, \bibinfo{address}{New York, NY, USA}.
\newblock
\urldef\tempurl%
\url{https://proceedings.neurips.cc/paper_files/paper/2019/file/a2186aa7c086b46ad4e8bf81e2a3a19b-Paper.pdf}
\showURL{%
\tempurl}


\bibitem[Meng et~al\mbox{.}(2023)]%
        {aux_rec_3}
\bibfield{author}{\bibinfo{person}{Chang Meng}, \bibinfo{person}{Ziqi Zhao}, \bibinfo{person}{Wei Guo}, \bibinfo{person}{Yingxue Zhang}, \bibinfo{person}{Haolun Wu}, \bibinfo{person}{Chen Gao}, \bibinfo{person}{Dong Li}, \bibinfo{person}{Xiu Li}, {and} \bibinfo{person}{Ruiming Tang}.} \bibinfo{year}{2023}\natexlab{}.
\newblock \showarticletitle{Coarse-to-Fine Knowledge-Enhanced Multi-Interest Learning Framework for Multi-Behavior Recommendation}.
\newblock \bibinfo{journal}{\emph{ACM Trans. Inf. Syst.}} \bibinfo{volume}{42}, \bibinfo{number}{1} (\bibinfo{year}{2023}), \bibinfo{pages}{1--27}.
\newblock
\href{https://doi.org/10.1145/3606369}{doi:\nolinkurl{10.1145/3606369}}


\bibitem[Pei et~al\mbox{.}(2025)]%
        {view_cl}
\bibfield{author}{\bibinfo{person}{Bo Pei}, \bibinfo{person}{Yingzheng Zhu}, \bibinfo{person}{Guangjin Wang}, \bibinfo{person}{Huajuan Duan}, \bibinfo{person}{Wenya Wu}, \bibinfo{person}{Fuyong Xu}, \bibinfo{person}{Yizhao Zhu}, \bibinfo{person}{Peiyu Liu}, {and} \bibinfo{person}{Ran Lu}.} \bibinfo{year}{2025}\natexlab{}.
\newblock \showarticletitle{Intent Contrastive Learning Based on Multi-view Augmentation for Sequential Recommendation}. In \bibinfo{booktitle}{\emph{Proceedings of the 31st International Conference on Computational Linguistics}} \emph{(\bibinfo{series}{COLING '25})}. \bibinfo{publisher}{Association for Computational Linguistics}, \bibinfo{pages}{3300--3309}.
\newblock
\urldef\tempurl%
\url{https://aclanthology.org/2025.coling-main.222/}
\showURL{%
\tempurl}


\bibitem[Qin et~al\mbox{.}(2023)]%
        {mclrec}
\bibfield{author}{\bibinfo{person}{Xiuyuan Qin}, \bibinfo{person}{Huanhuan Yuan}, \bibinfo{person}{Pengpeng Zhao}, \bibinfo{person}{Junhua Fang}, \bibinfo{person}{Fuzhen Zhuang}, \bibinfo{person}{Guanfeng Liu}, \bibinfo{person}{Yanchi Liu}, {and} \bibinfo{person}{Victor Sheng}.} \bibinfo{year}{2023}\natexlab{}.
\newblock \showarticletitle{Meta-optimized Contrastive Learning for Sequential Recommendation}. In \bibinfo{booktitle}{\emph{Proceedings of the 46th International ACM SIGIR Conference on Research and Development in Information Retrieval}} \emph{(\bibinfo{series}{SIGIR '23})}. \bibinfo{publisher}{Association for Computing Machinery}, \bibinfo{address}{New York, NY, USA}, \bibinfo{pages}{89--98}.
\newblock
\href{https://doi.org/10.1145/3539618.3591727}{doi:\nolinkurl{10.1145/3539618.3591727}}


\bibitem[Qin et~al\mbox{.}(2024)]%
        {icsrec}
\bibfield{author}{\bibinfo{person}{Xiuyuan Qin}, \bibinfo{person}{Huanhuan Yuan}, \bibinfo{person}{Pengpeng Zhao}, \bibinfo{person}{Guanfeng Liu}, \bibinfo{person}{Fuzhen Zhuang}, {and} \bibinfo{person}{Victor~S. Sheng}.} \bibinfo{year}{2024}\natexlab{}.
\newblock \showarticletitle{Intent Contrastive Learning with Cross Subsequences for Sequential Recommendation}. In \bibinfo{booktitle}{\emph{Proceedings of the 17th ACM International Conference on Web Search and Data Mining}} \emph{(\bibinfo{series}{WSDM '24})}. \bibinfo{publisher}{Association for Computing Machinery}, \bibinfo{address}{New York, NY, USA}, \bibinfo{pages}{548--556}.
\newblock
\href{https://doi.org/10.1145/3616855.3635773}{doi:\nolinkurl{10.1145/3616855.3635773}}


\bibitem[Qiu et~al\mbox{.}(2022)]%
        {duorec}
\bibfield{author}{\bibinfo{person}{Ruihong Qiu}, \bibinfo{person}{Zi Huang}, \bibinfo{person}{Hongzhi Yin}, {and} \bibinfo{person}{Zijian Wang}.} \bibinfo{year}{2022}\natexlab{}.
\newblock \showarticletitle{Contrastive Learning for Representation Degeneration Problem in Sequential Recommendation}. In \bibinfo{booktitle}{\emph{Proceedings of the Fifteenth ACM International Conference on Web Search and Data Mining}} \emph{(\bibinfo{series}{WSDM '22})}. \bibinfo{publisher}{Association for Computing Machinery}, \bibinfo{address}{New York, NY, USA}, \bibinfo{pages}{813--823}.
\newblock
\href{https://doi.org/10.1145/3488560.3498433}{doi:\nolinkurl{10.1145/3488560.3498433}}


\bibitem[Rendle et~al\mbox{.}(2010)]%
        {mc_2}
\bibfield{author}{\bibinfo{person}{Steffen Rendle}, \bibinfo{person}{Christoph Freudenthaler}, {and} \bibinfo{person}{Lars Schmidt-Thieme}.} \bibinfo{year}{2010}\natexlab{}.
\newblock \showarticletitle{Factorizing personalized Markov chains for next-basket recommendation}. In \bibinfo{booktitle}{\emph{Proceedings of the 19th International Conference on World Wide Web}} \emph{(\bibinfo{series}{WWW '10})}. \bibinfo{publisher}{Association for Computing Machinery}, \bibinfo{address}{New York, NY, USA}, \bibinfo{pages}{811--820}.
\newblock
\href{https://doi.org/10.1145/1772690.1772773}{doi:\nolinkurl{10.1145/1772690.1772773}}


\bibitem[Shao et~al\mbox{.}(2023)]%
        {intent_works_3}
\bibfield{author}{\bibinfo{person}{Weiqi Shao}, \bibinfo{person}{Xu Chen}, \bibinfo{person}{Jiashu Zhao}, \bibinfo{person}{Long Xia}, \bibinfo{person}{Jingsen Zhang}, {and} \bibinfo{person}{Dawei Yin}.} \bibinfo{year}{2023}\natexlab{}.
\newblock \showarticletitle{Sequential Recommendation with User Evolving Preference Decomposition}. In \bibinfo{booktitle}{\emph{Proceedings of the Annual International ACM SIGIR Conference on Research and Development in Information Retrieval in the Asia Pacific Region}} \emph{(\bibinfo{series}{SIGIR-AP '23})}. \bibinfo{publisher}{Association for Computing Machinery}, \bibinfo{address}{New York, NY, USA}, \bibinfo{pages}{253--263}.
\newblock
\href{https://doi.org/10.1145/3624918.3625312}{doi:\nolinkurl{10.1145/3624918.3625312}}


\bibitem[Sun et~al\mbox{.}(2019)]%
        {bert4rec}
\bibfield{author}{\bibinfo{person}{Fei Sun}, \bibinfo{person}{Jun Liu}, \bibinfo{person}{Jian Wu}, \bibinfo{person}{Changhua Pei}, \bibinfo{person}{Xiao Lin}, \bibinfo{person}{Wenwu Ou}, {and} \bibinfo{person}{Peng Jiang}.} \bibinfo{year}{2019}\natexlab{}.
\newblock \showarticletitle{BERT4Rec: Sequential Recommendation with Bidirectional Encoder Representations from Transformer}. In \bibinfo{booktitle}{\emph{Proceedings of the 28th ACM International Conference on Information and Knowledge Management}} \emph{(\bibinfo{series}{CIKM '19})}. \bibinfo{publisher}{Association for Computing Machinery}, \bibinfo{address}{New York, NY, USA}, \bibinfo{pages}{1441--1440}.
\newblock
\href{https://doi.org/10.1145/3357384.3357895}{doi:\nolinkurl{10.1145/3357384.3357895}}


\bibitem[Tian et~al\mbox{.}(2022)]%
        {mgnm}
\bibfield{author}{\bibinfo{person}{Yu Tian}, \bibinfo{person}{Jianxin Chang}, \bibinfo{person}{Yanan Niu}, \bibinfo{person}{Yang Song}, {and} \bibinfo{person}{Chenliang Li}.} \bibinfo{year}{2022}\natexlab{}.
\newblock \showarticletitle{When Multi-Level Meets Multi-Interest: A Multi-Grained Neural Model for Sequential Recommendation}. In \bibinfo{booktitle}{\emph{Proceedings of the 45th International ACM SIGIR Conference on Research and Development in Information Retrieval}} \emph{(\bibinfo{series}{SIGIR '22})}. \bibinfo{publisher}{Association for Computing Machinery}, \bibinfo{address}{New York, NY, USA}, \bibinfo{pages}{1632--1641}.
\newblock
\href{https://doi.org/10.1145/3477495.3532081}{doi:\nolinkurl{10.1145/3477495.3532081}}


\bibitem[Wang et~al\mbox{.}(2025b)]%
        {cross_domain_3}
\bibfield{author}{\bibinfo{person}{Maolin Wang}, \bibinfo{person}{Yutian Xiao}, \bibinfo{person}{Binhao Wang}, \bibinfo{person}{Sheng Zhang}, \bibinfo{person}{Shanshan Ye}, \bibinfo{person}{Wanyu Wang}, \bibinfo{person}{Hongzhi Yin}, \bibinfo{person}{Ruocheng Guo}, {and} \bibinfo{person}{Zenglin Xu}.} \bibinfo{year}{2025}\natexlab{b}.
\newblock \showarticletitle{FindRec: Stein-Guided Entropic Flow for Multi-Modal Sequential Recommendation}. In \bibinfo{booktitle}{\emph{Proceedings of the 31st ACM SIGKDD Conference on Knowledge Discovery and Data Mining V.2}} \emph{(\bibinfo{series}{KDD '25})}. \bibinfo{publisher}{Association for Computing Machinery}, \bibinfo{address}{New York, NY, USA}, \bibinfo{pages}{3008--3018}.
\newblock
\href{https://doi.org/10.1145/3711896.3736968}{doi:\nolinkurl{10.1145/3711896.3736968}}


\bibitem[Wang and Isola(2020)]%
        {draw_angles}
\bibfield{author}{\bibinfo{person}{Tongzhou Wang} {and} \bibinfo{person}{Phillip Isola}.} \bibinfo{year}{2020}\natexlab{}.
\newblock \showarticletitle{Understanding contrastive representation learning through alignment and uniformity on the hypersphere}. In \bibinfo{booktitle}{\emph{Proceedings of the 37th International Conference on Machine Learning}} \emph{(\bibinfo{series}{ICML '20})}. \bibinfo{publisher}{JMLR.org}, \bibinfo{pages}{9929--9939}.
\newblock


\bibitem[Wang et~al\mbox{.}(2025a)]%
        {ioclrec}
\bibfield{author}{\bibinfo{person}{Wuhong Wang}, \bibinfo{person}{Jianhui Ma}, \bibinfo{person}{Yuren Zhang}, \bibinfo{person}{Kai Zhang}, \bibinfo{person}{Junzhe Jiang}, \bibinfo{person}{Yihui Yang}, \bibinfo{person}{Yacong Zhou}, {and} \bibinfo{person}{Zheng Zhang}.} \bibinfo{year}{2025}\natexlab{a}.
\newblock \showarticletitle{Intent Oriented Contrastive Learning for Sequential Recommendation}.
\newblock \bibinfo{journal}{\emph{Proceedings of the AAAI Conference on Artificial Intelligence}} \bibinfo{volume}{39}, \bibinfo{number}{12} (\bibinfo{year}{2025}), \bibinfo{pages}{12748--12756}.
\newblock
\href{https://doi.org/10.1609/aaai.v39i12.33390}{doi:\nolinkurl{10.1609/aaai.v39i12.33390}}


\bibitem[Wu et~al\mbox{.}(2024)]%
        {gpr4dur}
\bibfield{author}{\bibinfo{person}{Haolun Wu}, \bibinfo{person}{Ofer Meshi}, \bibinfo{person}{Masrour Zoghi}, \bibinfo{person}{Fernando Diaz}, \bibinfo{person}{Xue Liu}, \bibinfo{person}{Craig Boutilier}, {and} \bibinfo{person}{Maryam Karimzadehgan}.} \bibinfo{year}{2024}\natexlab{}.
\newblock \showarticletitle{Density-based User Representation using Gaussian Process Regression for Multi-interest Personalized Retrieval}. In \bibinfo{booktitle}{\emph{Advances in Neural Information Processing Systems}} \emph{(\bibinfo{series}{NIPS '24})}. \bibinfo{publisher}{Curran Associates, Inc.}, \bibinfo{pages}{52568--52594}.
\newblock
\href{https://doi.org/10.52202/079017-1666}{doi:\nolinkurl{10.52202/079017-1666}}


\bibitem[Wu et~al\mbox{.}(2021)]%
        {cl_relate_work_2}
\bibfield{author}{\bibinfo{person}{Jiancan Wu}, \bibinfo{person}{Xiang Wang}, \bibinfo{person}{Fuli Feng}, \bibinfo{person}{Xiangnan He}, \bibinfo{person}{Liang Chen}, \bibinfo{person}{Jianxun Lian}, {and} \bibinfo{person}{Xing Xie}.} \bibinfo{year}{2021}\natexlab{}.
\newblock \showarticletitle{Self-supervised Graph Learning for Recommendation}. In \bibinfo{booktitle}{\emph{Proceedings of the 44th International ACM SIGIR Conference on Research and Development in Information Retrieval}} \emph{(\bibinfo{series}{SIGIR '21})}. \bibinfo{publisher}{Association for Computing Machinery}, \bibinfo{address}{New York, NY, USA}, \bibinfo{pages}{726--735}.
\newblock
\href{https://doi.org/10.1145/3404835.3462862}{doi:\nolinkurl{10.1145/3404835.3462862}}


\bibitem[Wu et~al\mbox{.}(2025)]%
        {intent_works_4}
\bibfield{author}{\bibinfo{person}{Wenhao Wu}, \bibinfo{person}{Xiaojie Li}, \bibinfo{person}{Lin Wang}, \bibinfo{person}{Jialiang Zhou}, \bibinfo{person}{Di Wu}, \bibinfo{person}{Qinye Xie}, \bibinfo{person}{Qingheng Zhang}, \bibinfo{person}{Yin Zhang}, \bibinfo{person}{Shuguang Han}, \bibinfo{person}{Fei Huang}, {and} \bibinfo{person}{Jufeng Chen}.} \bibinfo{year}{2025}\natexlab{}.
\newblock \showarticletitle{IU4Rec: Interest Unit-Based Product Organization and Recommendation for E-Commerce Platform}. In \bibinfo{booktitle}{\emph{Proceedings of the 31st ACM SIGKDD Conference on Knowledge Discovery and Data Mining V.2}} \emph{(\bibinfo{series}{KDD '25})}. \bibinfo{publisher}{Association for Computing Machinery}, \bibinfo{address}{New York, NY, USA}, \bibinfo{pages}{5039--5048}.
\newblock
\href{https://doi.org/10.1145/3711896.3737237}{doi:\nolinkurl{10.1145/3711896.3737237}}


\bibitem[Xiao et~al\mbox{.}(2020)]%
        {intent_attn_2}
\bibfield{author}{\bibinfo{person}{Zhibo Xiao}, \bibinfo{person}{Luwei Yang}, \bibinfo{person}{Wen Jiang}, \bibinfo{person}{Yi Wei}, \bibinfo{person}{Yi Hu}, {and} \bibinfo{person}{Hao Wang}.} \bibinfo{year}{2020}\natexlab{}.
\newblock \showarticletitle{Deep Multi-Interest Network for Click-through Rate Prediction}. In \bibinfo{booktitle}{\emph{Proceedings of the 29th ACM International Conference on Information \& Knowledge Management}} \emph{(\bibinfo{series}{CIKM '20})}. \bibinfo{publisher}{Association for Computing Machinery}, \bibinfo{address}{New York, NY, USA}, \bibinfo{pages}{2265--2268}.
\newblock
\href{https://doi.org/10.1145/3340531.3412092}{doi:\nolinkurl{10.1145/3340531.3412092}}


\bibitem[Xie et~al\mbox{.}(2022)]%
        {cl4srec}
\bibfield{author}{\bibinfo{person}{Xu Xie}, \bibinfo{person}{Fei Sun}, \bibinfo{person}{Zhaoyang Liu}, \bibinfo{person}{Shiwen Wu}, \bibinfo{person}{Jinyang Gao}, \bibinfo{person}{Jiandong Zhang}, \bibinfo{person}{Bolin Ding}, {and} \bibinfo{person}{Bin Cui}.} \bibinfo{year}{2022}\natexlab{}.
\newblock \showarticletitle{Contrastive Learning for Sequential Recommendation}. In \bibinfo{booktitle}{\emph{2022 IEEE 38th International Conference on Data Engineering (ICDE)}} \emph{(\bibinfo{series}{ICDE '22})}. \bibinfo{publisher}{IEEE}, \bibinfo{pages}{1259--1273}.
\newblock
\href{https://doi.org/10.1109/ICDE53745.2022.00099}{doi:\nolinkurl{10.1109/ICDE53745.2022.00099}}


\bibitem[Xie et~al\mbox{.}(2023)]%
        {remi}
\bibfield{author}{\bibinfo{person}{Yueqi Xie}, \bibinfo{person}{Jingqi Gao}, \bibinfo{person}{Peilin Zhou}, \bibinfo{person}{Qichen Ye}, \bibinfo{person}{Yining Hua}, \bibinfo{person}{Jae~Boum Kim}, \bibinfo{person}{Fangzhao Wu}, {and} \bibinfo{person}{Sunghun Kim}.} \bibinfo{year}{2023}\natexlab{}.
\newblock \showarticletitle{Rethinking Multi-Interest Learning for Candidate Matching in Recommender Systems}. In \bibinfo{booktitle}{\emph{Proceedings of the 17th ACM Conference on Recommender Systems}} \emph{(\bibinfo{series}{RecSys '23})}. \bibinfo{publisher}{Association for Computing Machinery}, \bibinfo{address}{New York, NY, USA}, \bibinfo{pages}{283--293}.
\newblock
\href{https://doi.org/10.1145/3604915.3608766}{doi:\nolinkurl{10.1145/3604915.3608766}}


\bibitem[Ying et~al\mbox{.}(2025)]%
        {cross_domain_1}
\bibfield{author}{\bibinfo{person}{Xiang Ying}, \bibinfo{person}{Rui Ding}, \bibinfo{person}{Yue Zhao}, \bibinfo{person}{Mei Yu}, {and} \bibinfo{person}{Mankun Zhao}.} \bibinfo{year}{2025}\natexlab{}.
\newblock \showarticletitle{DPT: Dynamic Preference Transfer for Cross-Domain Sequential Recommendation}. In \bibinfo{booktitle}{\emph{Proceedings of the 34th ACM International Conference on Information and Knowledge Management}} \emph{(\bibinfo{series}{CIKM '25})}. \bibinfo{publisher}{Association for Computing Machinery}, \bibinfo{address}{New York, NY, USA}, \bibinfo{pages}{3909--3919}.
\newblock
\href{https://doi.org/10.1145/3746252.3761075}{doi:\nolinkurl{10.1145/3746252.3761075}}


\bibitem[Yu et~al\mbox{.}(2022)]%
        {simgcl}
\bibfield{author}{\bibinfo{person}{Junliang Yu}, \bibinfo{person}{Hongzhi Yin}, \bibinfo{person}{Xin Xia}, \bibinfo{person}{Tong Chen}, \bibinfo{person}{Lizhen Cui}, {and} \bibinfo{person}{Quoc Viet~Hung Nguyen}.} \bibinfo{year}{2022}\natexlab{}.
\newblock \showarticletitle{Are Graph Augmentations Necessary? Simple Graph Contrastive Learning for Recommendation}. In \bibinfo{booktitle}{\emph{Proceedings of the 45th International ACM SIGIR Conference on Research and Development in Information Retrieval}} \emph{(\bibinfo{series}{SIGIR '22})}. \bibinfo{publisher}{Association for Computing Machinery}, \bibinfo{address}{New York, NY, USA}, \bibinfo{pages}{1294--1303}.
\newblock
\href{https://doi.org/10.1145/3477495.3531937}{doi:\nolinkurl{10.1145/3477495.3531937}}


\bibitem[Yuan et~al\mbox{.}(2025)]%
        {srec_survey_3}
\bibfield{author}{\bibinfo{person}{Jun Yuan}, \bibinfo{person}{Guohao Cai}, {and} \bibinfo{person}{Zhenhua Dong}.} \bibinfo{year}{2025}\natexlab{}.
\newblock \showarticletitle{A Contextual-Aware Position Encoding for Sequential Recommendation}. In \bibinfo{booktitle}{\emph{Companion Proceedings of the ACM on Web Conference 2025}} \emph{(\bibinfo{series}{WWW '25})}. \bibinfo{publisher}{Association for Computing Machinery}, \bibinfo{address}{New York, NY, USA}, \bibinfo{pages}{577--585}.
\newblock
\href{https://doi.org/10.1145/3701716.3715206}{doi:\nolinkurl{10.1145/3701716.3715206}}


\bibitem[Zhang et~al\mbox{.}(2022)]%
        {re4}
\bibfield{author}{\bibinfo{person}{Shengyu Zhang}, \bibinfo{person}{Lingxiao Yang}, \bibinfo{person}{Dong Yao}, \bibinfo{person}{Yujie Lu}, \bibinfo{person}{Fuli Feng}, \bibinfo{person}{Zhou Zhao}, \bibinfo{person}{Tat-seng Chua}, {and} \bibinfo{person}{Fei Wu}.} \bibinfo{year}{2022}\natexlab{}.
\newblock \showarticletitle{Re4: Learning to Re-contrast, Re-attend, Re-construct for Multi-interest Recommendation}. In \bibinfo{booktitle}{\emph{Proceedings of the ACM Web Conference 2022}} \emph{(\bibinfo{series}{WWW '22})}. \bibinfo{publisher}{Association for Computing Machinery}, \bibinfo{address}{New York, NY, USA}, \bibinfo{pages}{2216--2226}.
\newblock
\href{https://doi.org/10.1145/3485447.3512094}{doi:\nolinkurl{10.1145/3485447.3512094}}


\bibitem[Zhang et~al\mbox{.}(2024)]%
        {soft_cl}
\bibfield{author}{\bibinfo{person}{Yabin Zhang}, \bibinfo{person}{Zhenlei Wang}, \bibinfo{person}{Wenhui Yu}, \bibinfo{person}{Lantao Hu}, \bibinfo{person}{Peng Jiang}, \bibinfo{person}{Kun Gai}, {and} \bibinfo{person}{Xu Chen}.} \bibinfo{year}{2024}\natexlab{}.
\newblock \showarticletitle{Soft Contrastive Sequential Recommendation}.
\newblock \bibinfo{journal}{\emph{ACM Trans. Inf. Syst.}} \bibinfo{volume}{42}, \bibinfo{number}{6} (\bibinfo{year}{2024}), \bibinfo{pages}{1--28}.
\newblock
\href{https://doi.org/10.1145/3665325}{doi:\nolinkurl{10.1145/3665325}}


\bibitem[Zhou et~al\mbox{.}(2020)]%
        {s3rec}
\bibfield{author}{\bibinfo{person}{Kun Zhou}, \bibinfo{person}{Hui Wang}, \bibinfo{person}{Wayne~Xin Zhao}, \bibinfo{person}{Yutao Zhu}, \bibinfo{person}{Sirui Wang}, \bibinfo{person}{Fuzheng Zhang}, \bibinfo{person}{Zhongyuan Wang}, {and} \bibinfo{person}{Ji-Rong Wen}.} \bibinfo{year}{2020}\natexlab{}.
\newblock \showarticletitle{S3-Rec: Self-Supervised Learning for Sequential Recommendation with Mutual Information Maximization}. In \bibinfo{booktitle}{\emph{Proceedings of the 29th ACM International Conference on Information \& Knowledge Management}} \emph{(\bibinfo{series}{CIKM '20})}. \bibinfo{publisher}{Association for Computing Machinery}, \bibinfo{address}{New York, NY, USA}, \bibinfo{pages}{1893--1902}.
\newblock
\href{https://doi.org/10.1145/3340531.3411954}{doi:\nolinkurl{10.1145/3340531.3411954}}


\bibitem[Zhu et~al\mbox{.}(2021)]%
        {gnn_based_1}
\bibfield{author}{\bibinfo{person}{Yanqiao Zhu}, \bibinfo{person}{Yichen Xu}, \bibinfo{person}{Feng Yu}, \bibinfo{person}{Qiang Liu}, \bibinfo{person}{Shu Wu}, {and} \bibinfo{person}{Liang Wang}.} \bibinfo{year}{2021}\natexlab{}.
\newblock \showarticletitle{Graph Contrastive Learning with Adaptive Augmentation}. In \bibinfo{booktitle}{\emph{Proceedings of the Web Conference 2021}} \emph{(\bibinfo{series}{WWW '21})}. \bibinfo{publisher}{Association for Computing Machinery}, \bibinfo{address}{New York, NY, USA}, \bibinfo{pages}{2069--2080}.
\newblock
\href{https://doi.org/10.1145/3442381.3449802}{doi:\nolinkurl{10.1145/3442381.3449802}}


\end{thebibliography}

\clearpage

\appendix

\section{RELATED WORK}
\label{sec:related_work}

\subsection{Sequential Recommendation}

Early studies on sequential recommendation primarily relied on Markov Chains (MC) to model local transition patterns in user behavior~\cite{mc_1,mc_2}. Subsequent works introduced deep neural architectures for more expressive sequence modeling, with representative approaches including recurrent neural networks (RNNs)~\cite{rnn}, memory networks~\cite{memory_1}, attentional mechanism~\cite{sasrec,bert4rec}, and graph neural networks (GNNs)~\cite{gnn_based_1}. Among these, self-attention-based models, such as \emph{SASRec}~\cite{sasrec} and \emph{Bert4Rec}~\cite{bert4rec}, have demonstrated strong empirical performance by adaptively weighting historical interactions and effectively capturing long-range temporal dependencies. Complementary to architectural advances, another line of research~\cite{aux_signs_1,aux_signs_2} incorporates auxiliary signals—such as item attributes, categorical information, and external knowledge—to alleviate data sparsity and enrich user representations. For instance, \emph{FDSA}~\cite{fdsa} fuses heterogeneous item attributes via feature-level self-attention. Recent works~\cite{cross_domain_1,cross_domain_2,cross_domain_3} further exploit cross-domain or multi-modal information to enrich sequential modeling. Despite their effectiveness, these methods typically encode user preferences into a single latent representation. This design limits their ability to disentangle diverse user interests and capture the multi-faceted nature of sequential behaviors.

\subsection{Multi-intent Recommendation}

Multi-intent recommendation has emerged as an effective paradigm for modeling diverse user preferences, with existing methods differing in how latent intents are inferred and utilized. \textbf{\emph{Auxiliary knowledge-based methods}}~\cite{aux_rec_1,aux_rec_2,aux_rec_3,aux_rec_4} leverage various side information—such as user profiles, item attributes, multi-type behaviors, knowledge graphs, user groups, or multimodal signals—to guide intent disentanglement. Yet, these methods heavily depend on high-quality auxiliary information, which may be unavailable or noisy in many real-world scenarios. In contrast, \textbf{\emph{interaction-based methods}} infer multiple user intents solely from interaction histories, typically using attention mechanisms~\cite{simrec,intent_attn_1,intent_attn_2} or capsule networks~\cite{mind,comirec,mgnm,remi}. Representative works employ dynamic routing to extract multiple intent vectors from engaged items~\cite{mind}, or integrate sequential capsule networks with graph-based aggregation to capture multi-level preferences~\cite{mgnm}. While effective in modeling preference diversity, these methods mainly focus on intra-sequence disentanglement and overlook shared intent patterns across users, making them sensitive to sparsity and noise in individual sequences. Some recent works~\cite{iclrec,icsrec,elcrec} further incorporate clustering or contrastive learning to derive latent global intent prototypes. Nevertheless, intent signals in these models are generally used as auxiliary supervision during training, which limits the effective exploitation of global intent priors for robust and discriminative sequence modeling.

\subsection{Contrastive Learning for Recommendation}

Contrastive learning has been increasingly adopted in sequential recommendation as a self-supervised learning (SSL) paradigm improve representation quality and robustness under sparse or noisy data~\cite{cl_survey}. Early efforts such as $S^3$-Rec~\cite{s3rec} introduce contrastive objectives via multiple auxiliary SSL tasks. Subsequent works (e.g., \emph{CL4SRec}~\cite{cl4srec}, \emph{CoSeRec}~\cite{coserec}, \emph{TiCoSeRec}~\cite{TiCoSeRec}, \emph{ELCRec}~\cite{elcrec}, \emph{MCLRec}~\cite{mclrec}, \emph{SoftCSR}~\cite{soft_cl}, and \emph{ICMA}~\cite{view_cl}) primarily construct contrastive views using \textbf{\emph{sequence-level random augmentations}}. While increasing view diversity, these augmentations may disrupt temporal order or long-range dependencies, encouraging invariances that conflict with the underlying behavioral semantics. In parallel, \textbf{\emph{GNN-based approaches}}~\cite{cl_relate_work_2,gnn_based_1} explore SSL over interaction graphs to improve structural representation learning. 
Some recent approaches including \emph{ICLRec}~\cite{iclrec}, \emph{ICSRec}~\cite{icsrec} and \emph{IOCRec}~\cite{ioclrec} incorporate \textbf{\emph{intent-aware signals}} to guide contrastive learning. They align sequences based on latent intent distributions. However, these methods rely on aggregated intent prototypes, which may suppress instance-specific variations and yield relatively homogeneous contrastive targets, thereby limiting discriminative learning. Furthermore, methods like \emph{DuoRec}~\cite{duorec} and \emph{ICSRec}~\cite{icsrec} assume that sequences sharing the same target item have similar intents, and thus treat them as positive pairs in the contrastive objective. This assumption ignores that the same item can correspond to different underlying user intents (e.g., entertainment, study, or work; see Fig.~\ref{fig:intent_augs}). Consequently, semantically distinct sequences may be forced to align, potentially degrading representation discriminability.

\section{Overall Training Procedure of \emph{BIPCL}}
\label{procedure_code}

Algorithm~\ref{alg:BIPCL} summarizes the overall training procedure of \emph{BIPCL}. In brief, given user interaction sequences and the global item co-occurrence graph, the model jointly optimizes the recommendation objective and a multi-level contrastive learning objective in an end-to-end manner. Specifically, \emph{BIPCL} first performs lightweight graph propagation over the item co-occurrence graph to obtain structurally informed item embeddings. These embeddings are then processed by the intent-enhanced item encoder and sequence encoder, producing item and sequence representations for next-item prediction. During this process, intent embeddings serve as intermediate signals to enhance representation learning by capturing collective behavioral patterns. To construct semantically consistent yet discriminative contrastive views, \emph{BIPCL} injects bounded, direction-aware perturbations directly into structural item embeddings. The perturbed embeddings are propagated through both intent-enhanced encoders, and contrastive alignment is enforced at both the interaction level and the intent level using the InfoNCE objective. Finally, the recommendation and contrastive losses are jointly optimized. Early stopping is applied based on validation performance to prevent overfitting.

\begin{algorithm}[t]
\caption{Training Procedure of \emph{BIPCL}}
\label{alg:BIPCL}

\tcp{%
$\mathcal{F}_{\mathrm{item}}$: Intent-aware Item Representation Learning module (Sec.~\ref{item_encoder})\\
$\mathcal{F}_{\mathrm{seq}}$: Intent-aware Sequential Representation Learning module (Sec.~\ref{seq_encoder})
}

\KwIn{
Training instances $\mathcal{D}_{\mathrm{train}}$
(consisting of user sequences and ground-truth next-item labels); \\
validation instances $\mathcal{D}_{\mathrm{val}}$; 
item co-occurrence graph $\textbf{A}$; \\
model parameters $\Theta$;
noise scale $\varepsilon$;
contrastive weight $\lambda$
}
\KwOut{Optimized model parameters $\Theta$}

Initialize $\Theta$\;

\While{early stopping criterion not met on $\mathcal{D}_{\mathrm{val}}$}{
    Sample a mini-batch of training instances
    $\mathcal{B} = \{(\mathcal{S}^u, v^+)\} \subset \mathcal{D}_{\mathrm{train}}$\;
    
    \tcp{Structural Item Encoding}
    Compute high-order structural item embeddings
    $\mathbf{R}^{L}$ via graph propagation on $\textbf{A}$\;
    
    \tcp{Recommendation Objective}
    Obtain item representations $\mathbf{h}^{(v)}$ via
    $\mathcal{F}_{\mathrm{item}}(\mathbf{R}^{L})$\;
    Obtain user sequence representations $\mathbf{h}^{(u)}$ via $\mathcal{F}_{\mathrm{seq}}(\mathbf{R}^{L})$\;
    Compute next-item prediction loss $\mathcal{L}_{\mathrm{rec}}$ (Eq.~\ref{eq:rec_loss}) on $\mathcal{B}$\;
    
    \tcp{Perturbation-based View Generation}
    Sample $\boldsymbol{\xi}^{(1)}, \boldsymbol{\xi}^{(2)} \sim \mathcal{N}(\mathbf{0}, \mathbf{I})$\;
    Generate perturbed structural views
    $\widetilde{\mathbf{R}}^{(k)} =
    \mathbf{R}^{L}
    + \varepsilon \cdot \mathrm{sign}(\mathbf{R}^{L})
    \odot \mathrm{norm}(\boldsymbol{\xi}^{(k)})$,
    $k \in \{1,2\}$\;
    
    \tcp{Intent-aware Encoding under Perturbations}
    Obtain perturbed item representations and item intents $\left \{ \mathbf{h}^{(v,k)}, \mathbf{z}^{(v,k)} \right \}$ via $
    \mathcal{F}_{\mathrm{item}}(\widetilde{\mathbf{R}}^{(k)})$\;
    Obtain perturbed user sequence representations and user sequence intents $\left \{ \mathbf{h}^{(u,k)}, \mathbf{z}^{(u,k)} \right \}$ via $
    \mathcal{F}_{\mathrm{seq}}(\widetilde{\mathbf{R}}^{(k)})$,
    $k \in \{1,2\}$\;
    
    \tcp{Multi-level Contrastive Alignment (Batch-wise)}
    Compute contrastive loss $\mathcal{L}_{\mathrm{CL}}$ (Eq.~\ref{eq:cl_loss}) via InfoNCE over
    $\{(\mathbf{h}^{(v,1)}, \mathbf{h}^{(v,2)}),
      (\mathbf{z}^{(v,1)}, \mathbf{z}^{(v,2)}),
      (\mathbf{h}^{(u,1)}, \mathbf{h}^{(u,2)}),
      (\mathbf{z}^{(u,1)}, \mathbf{z}^{(u,2)})\}$,
    using other instances in $\mathcal{B}$ as negatives\;
    
    \tcp{Joint Optimization}
    Update $\Theta$ by minimizing
    $\mathcal{L} = \mathcal{L}_{\mathrm{rec}} + \lambda \mathcal{L}_{\mathrm{CL}}$\;
    
    \tcp{Validation and Early Stopping}
    Evaluate next-item prediction performance on
    $\mathcal{D}_{\mathrm{val}}$ and update early stopping status\;
}
\end{algorithm}

\section{Runtime and Memory Consumption Analysis}
\label{append_time}

We evaluate the computational efficiency of \emph{BIPCL} in terms of both runtime and memory usage, comparing it with representative multi-intent SR baselines. All experiments are performed on a server with 16 vCPU Intel(R) Xeon(R) Platinum 8481C processors and a single NVIDIA RTX 4090D GPU with 24GB memory. To ensure fair comparison, the batch size is fixed to 256 for all models.

\textbf{\emph{Runtime.}} 
We measure the average training time per batch and the inference time over the entire test set. Specifically, training time is averaged over the first 500 batches. Table~\ref{tab:time_comparison} summarizes the runtime results. Compared with \emph{SimRec}, \emph{BIPCL} incurs a modest increase in training time due to the additional intent-enhanced encoding and multi-level contrastive objectives. Nevertheless, it remains faster than the next strongest baseline, \emph{DisMIR}. Considering the consistent performance improvements achieved by \emph{BIPCL}, this slight increase in training cost is well within practical limits. Notably, by caching enhanced item representations and avoiding explicit multi-intent matching during inference, \emph{BIPCL} achieves superior inference efficiency compared to all baselines.

\begin{figure}[t]
    \centering
    \includegraphics[width=\columnwidth]{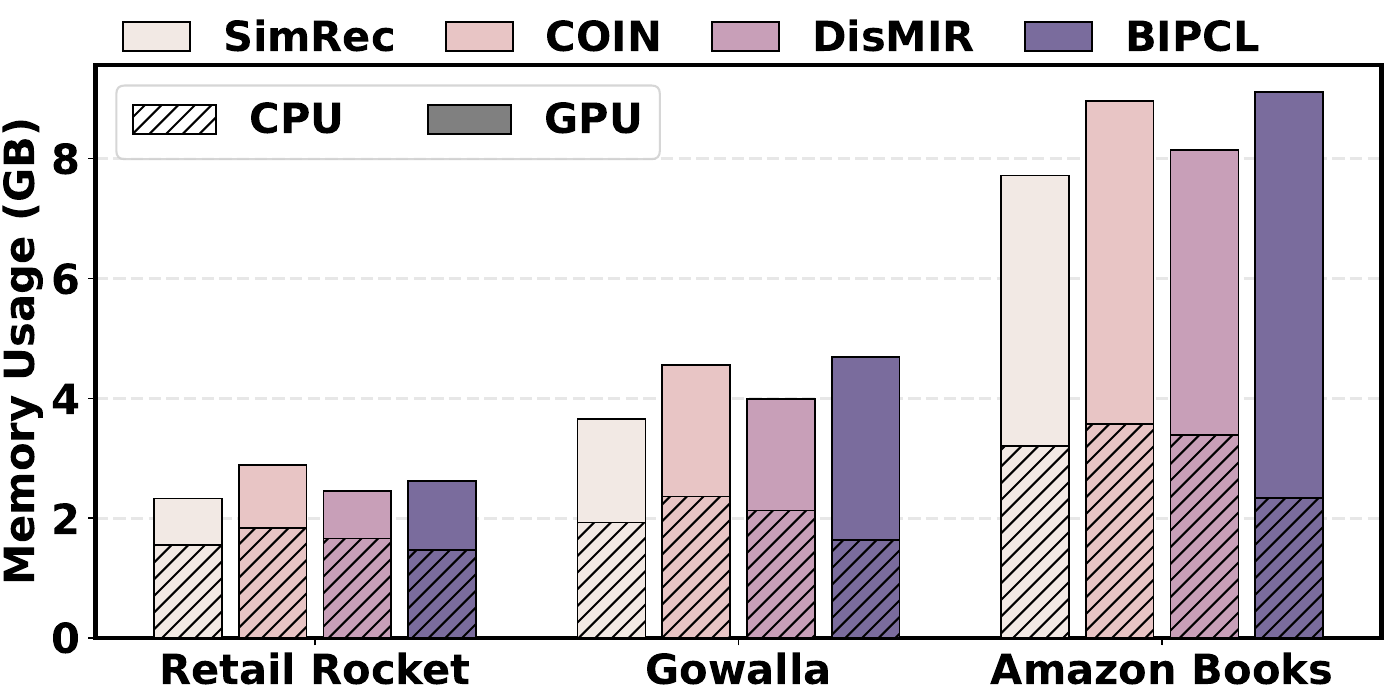}
    \caption{CPU and GPU memory consumption of \emph{BIPCL} and baseline methods on three large-scale datasets.}
    \label{fig:memory_consumption}
\end{figure}

\textbf{\emph{Memory Consumption.}} 
Figure~\ref{fig:memory_consumption} reports CPU and GPU memory usage over the complete training pipeline. \emph{BIPCL} exhibits comparable overall memory usage to existing baselines. To accelerate training, our implementation allocates most computations to the GPU, resulting in slightly higher GPU memory usage. Meanwhile, this design substantially reduces CPU memory consumption, leading to an efficient balance between GPU and CPU resources. These results demonstrate that \emph{BIPCL} achieves its performance improvements without introducing excessive memory overhead.

\begin{table}[t]
\centering
\caption{Training time per batch and testing time (s).}
\label{tab:time_comparison}
\setlength{\tabcolsep}{6pt}
\resizebox{\columnwidth}{!}{
\begin{tabular}{lcccccc}
\toprule
\multirow{2}{*}{Models}
& \multicolumn{2}{c}{Retail Rocket}
& \multicolumn{2}{c}{Gowalla}
& \multicolumn{2}{c}{Amazon Books} \\
\cmidrule(lr){2-3}
\cmidrule(lr){4-5}
\cmidrule(lr){6-7}
& train & test
& train & test
& train & test \\
\midrule
REMI      
& \textbf{\textcolor{deepurple}{0.0142}} & 0.9773 
& \textbf{\textcolor{deepurple}{0.0418}} & 1.9365 
& \textbf{\textcolor{deepurple}{0.1124}} & 10.9235 \\
SimRec     & 0.0160 & 0.7151 & 0.0456 & 1.4927 & 0.1378 & 9.7941 \\
COIN     & 0.0501 & 1.2635 & 0.0914 & 2.1720 & 0.2266 & 14.8637 \\
DisMIR     & 0.0994 & 1.4870 & 0.1647 & 2.7689 & 1.1613 & 17.4763 \\
BIPCL     
& 0.0732 
& \textbf{\textcolor{deepurple}{0.3724}} 
& 0.1213 
& \textbf{\textcolor{deepurple}{0.6216}} 
& 0.3049 
& \textbf{\textcolor{deepurple}{2.6920}} \\
\bottomrule
\end{tabular}
}
\end{table}

\begin{table}[t]
  \caption{Ablation results of \emph{BIPCL} on different datasets.}
  \label{tab:append_abs_study}
  \centering
  \setlength{\tabcolsep}{3pt}
  \resizebox{\columnwidth}{!}{
  \begin{tabular}{lcccccc}
    \toprule
    \multirow{2}{*}{Model} 
    & \multicolumn{3}{c}{Gowalla} 
    & \multicolumn{3}{c}{Amazon Books} \\
    \cmidrule(lr){2-4} \cmidrule(lr){5-7}
    & R@20 & ND@20 & HR@20 & R@20 & ND@20 & HR@20 \\
    \midrule
    w/o Intent       
    & 0.1304 & 0.1912 & 0.4051  
    & 0.0726 & 0.0614 & 0.1552 \\
    w/o Gating       
    & 0.1476 & 0.2138 & 0.4437 
    & 0.0874 & 0.0768 & 0.1759 \\
    w/o Pooling      
    & 0.1416 & 0.2030 & 0.4269 
    & 0.0803 & 0.0706 & 0.1663 \\
    Graph Aug        
    & 0.1447 & 0.2095 & 0.4346  
    & 0.0854 & 0.0760 & 0.1741 \\
    Seq Aug          
    & 0.1432 & 0.2043 & 0.4315
    & 0.0837 & 0.0732 & 0.1715 \\
    w/o CL    
    & 0.1393 & 0.2022 & 0.4304 
    & 0.0796 & 0.0727 & 0.1688 \\
    w/o Final-CL     
    & 0.1422 & 0.2076 & 0.4352  
    & 0.0830 & 0.0745 & 0.1713 \\
    w/o Intent-CL    
    & 0.1479 & 0.2124 & 0.4420
    & 0.0884 & 0.0793 & 0.1786 \\
    \midrule
    \textbf{BIPCL}    
    & \textbf{\textcolor{deepurple}{0.1506}}  
    & \textbf{\textcolor{deepurple}{0.2164}} 
    & \textbf{\textcolor{deepurple}{0.4485}} 
    & \textbf{\textcolor{deepurple}{0.0901}} 
    & \textbf{\textcolor{deepurple}{0.0815}} 
    & \textbf{\textcolor{deepurple}{0.1824}}  \\
    \bottomrule
  \end{tabular}
  }
\end{table}

\begin{table*}[t]
  \caption{Performance of \emph{BIPCL} under different sequential encoder configurations.}
  \label{tab:params_analysis_append}
  \centering
  \setlength{\tabcolsep}{3pt}
  \resizebox{\textwidth}{!}{
  \begin{tabular}{ccccccccccccccccc}
    \toprule
    \multirow{2}{*}{\#Blocks}
    & \multirow{2}{*}{\#Heads}
    & \multicolumn{3}{c}{Beauty} 
    & \multicolumn{3}{c}{Yelp} 
    & \multicolumn{3}{c}{Retail Rocket}
    & \multicolumn{3}{c}{Gowalla}  
    & \multicolumn{3}{c}{Amazon Books}  \\
    \cmidrule(lr){3-5} \cmidrule(lr){6-8} \cmidrule(lr){9-11}
    \cmidrule(lr){12-14}  \cmidrule(lr){15-17}
    & 
    & R@20 & ND@20 & HR@20
    & R@20 & ND@20 & HR@20
    & R@20 & ND@20 & HR@20 
    & R@20 & ND@20 & HR@20
    & R@20 & ND@20 & HR@20   \\
    \midrule
    \multirow{3}{*}{1}
    & 2
    & 0.1247 & 0.0892 & 0.2035
    & 0.1182 & 0.0918 & 0.2296
    & 0.2513 & 0.1618 & 0.3714 
    & 0.1407 & 0.2063 & 0.4369
    & 0.0776 & 0.0695 & 0.2386 \\
    & 4
    & \textbf{\textcolor{deepurple}{0.1277}} 
    & 0.0909
    & \textbf{\textcolor{deepurple}{0.2074}}
    & \textbf{\textcolor{deepurple}{0.1228}} 
    & \textbf{\textcolor{deepurple}{0.0963}} 
    & \textbf{\textcolor{deepurple}{0.2343}}  
    & 0.2559 & 0.1648 & 0.3782 
    & 0.1451 & 0.2104 & 0.4393 
    & 0.0824 & 0.0731 & 0.2470  \\
    & 8
    & 0.1256 & 0.0899 & 0.2043
    & 0.1202 & 0.0937 & 0.2320
    & 0.2535 & 0.1633 & 0.3756   
    & 0.1464 & 0.2120 & 0.4412  
    & 0.0837 & 0.0742 & 0.2455  \\
    \hline
    \multirow{3}{*}{2}
    & 2
    & 0.1204 & 0.0889 & 0.2005 
    & 0.1158 & 0.0907 & 0.2242 
    & 0.2555 & 0.1642 & 0.3770   
    & 0.1470 & 0.2122 & 0.4415 
    & 0.0874 & 0.0796 & 0.2602  \\
    & 4
    & 0.1239 & \textbf{\textcolor{deepurple}{0.0915}} & 0.2039
    & 0.1190 & 0.0933 & 0.2281 
    & \textbf{\textcolor{deepurple}{0.2580}}
    & \textbf{\textcolor{deepurple}{0.1649}} 
    & \textbf{\textcolor{deepurple}{0.3788}} 
    & \textbf{\textcolor{deepurple}{0.1506}} 
    & 0.2164  
    & \textbf{\textcolor{deepurple}{0.4485}}  
    & 0.0901
    & 0.0815
    & 0.2638  
    \\
    & 8
    & 0.1222 & 0.0903 & 0.2017 
    & 0.1179 & 0.0929 & 0.2260  
    & 0.2568 & 0.1642 
    & \textbf{\textcolor{deepurple}{0.3788}}  
    & 0.1494 & 0.2148 & 0.4473
    & \textbf{\textcolor{deepurple}{0.0912}} 
    & 0.0801 & 0.2633  \\
    \hline
    \multirow{3}{*}{3}
    & 2
    & 0.1187 & 0.0862 & 0.1992 
    & 0.1128 & 0.0896 & 0.2218  
    & 0.2537 & 0.1609 & 0.3738  
    & 0.1432 & 0.2169 & 0.4416 
    & 0.0864 & 0.0795 & 0.2588 \\
    & 4
    & 0.1220 & 0.0884 & 0.2013
    & 0.1164 & 0.0922 & 0.2274 
    & 0.2571 
    & 0.1633 
    & 0.3776   
    & 0.1487 
    & \textbf{\textcolor{deepurple}{0.2212}} 
    & 0.4461   
    & 0.0886 
    & \textbf{\textcolor{deepurple}{0.0839}} 
    & \textbf{\textcolor{deepurple}{0.2645}}  \\
    & 8
    & 0.1212 & 0.0879 & 0.2010 
    & 0.1142 & 0.0911 & 0.2235  
    & 0.2553 & 0.1629 & 0.3762  
    & 0.1466 & 0.2202 & 0.4438 
    & 0.0892
    & 0.0820
    & 0.2614  \\
    \bottomrule
  \end{tabular}
  }
\end{table*}

\begin{figure}[t]
    \centering
    \includegraphics[width=\columnwidth]{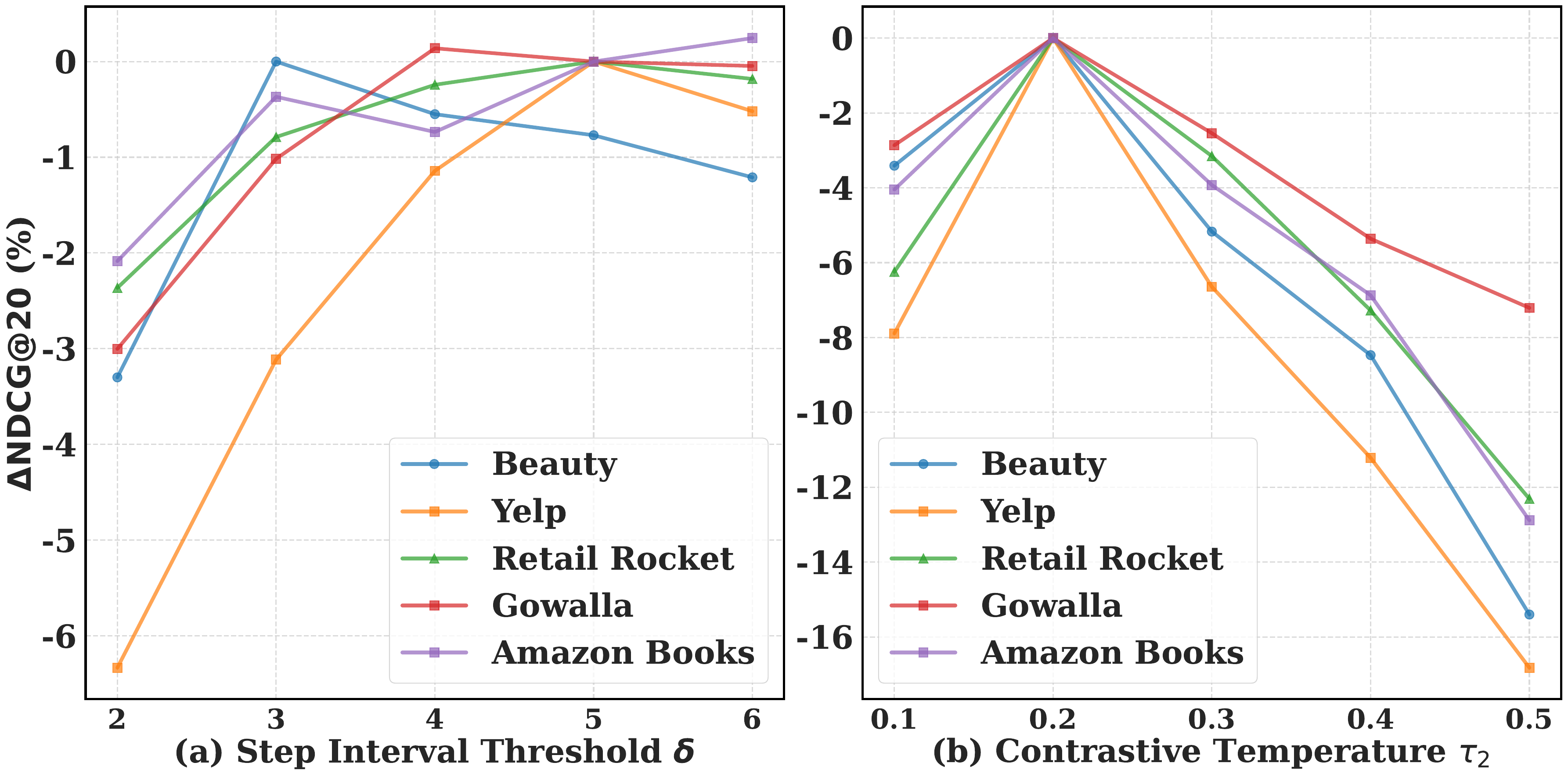}
    \caption{Hyperparameter sensitivity of \emph{BIPCL}.}
    \label{fig:params_analysis_append}
\end{figure}

\section{Additional Ablation Studies}
\label{append_abs}

We further report ablation results of eight \emph{BIPCL} variants on the Gowalla and Amazon Books datasets (Table~\ref{tab:append_abs_study}). 
The overall trends are consistent with the main results presented in Sec.~\ref{abs_study} and support the following observations. \textbf{\emph{(A) Intent Modeling and Fusion.}} 
Removing the intent enhancement module (\textbf{\emph{w/o Intent}}) or replacing the sequence encoder with an average-pooled representation (\textbf{\emph{w/o Pooling}}) substantially degrades performance. 
These results demonstrate the importance of explicit multi-intent modeling and long-term sequential dependency capture in \emph{BIPCL}. 
In contrast, replacing the gated fusion mechanism with direct aggregation (\textbf{\emph{w/o Gating}}) causes a relatively moderate performance drop, indicating that gated fusion primarily serves as a refinement rather than a dominant contributor. 
\textbf{\emph{(B) Augmentation Strategies.}} 
Alternative perturbation schemes—graph-level (\textbf{\emph{Graph Aug}}) or sequence-level (\textbf{\emph{Seq Aug}})—consistently underperform the proposed embedding-level perturbation scheme. Moreover, these strategies introduce significant computational overhead, increasing both training time and memory usage, which limits their practical applicability. \textbf{\emph{(C) Multi-level Contrastive Learning.}} 
Removing the multi-level contrastive learning objective significantly reduces performance, highlighting its critical role in representation learning. Specifically, eliminating interaction-level contrastive alignment (\textbf{\emph{w/o Final-CL}}) has a larger impact than removing intent-level alignment (\textbf{\emph{w/o Intent-CL}}), indicating that fine-grained interaction consistency provides stronger supervision than high-level intent alignment alone. 
Overall, these extended ablation studies confirm that each component of \emph{BIPCL} contributes positively, and that the proposed design choices work synergistically to achieve the observed performance gains.

\section{Additional Hyperparameter Analysis}
\label{sec:appendix_params}

We further examine the sensitivity of \emph{BIPCL} to key hyperparameters, including the configuration of the sequential encoder $\mathrm{TransEnc}(\cdot)$, the construction of the item co-occurrence graph (step interval threshold $\delta$), and the temperature parameters $\tau_1$ and $\tau_2$.

\textbf{\emph{Sequential Encoder Configuration.}}
Table~\ref{tab:params_analysis_append} summarizes the performance of \emph{BIPCL} under different numbers of self-attention blocks and attention heads. 
\textbf{\emph{Encoder Depth.}} Overall, \emph{BIPCL} is robust to the number of self-attention blocks, provided that sufficient model capacity is ensured. On small datasets (Beauty and Yelp), a single block achieves the best performance; increasing depth leads to slight degradation, suggesting limited benefits from additional capacity. In contrast, on larger datasets (Gowalla, and Amazon Books), using only one block results in insufficient modeling capacity and noticeably worse performance. Increasing depth to two blocks substantially improves results, after which further deepening yields marginal gains and stable results. \textbf{\emph{Number of Attention Heads.}} \emph{BIPCL} consistently performs best with four heads across datasets and encoder depths. Using fewer heads limits the expressiveness of multi-head attention and weakens the modeling of diverse sequential patterns. Employing too many heads, however, fragments the representation space and reduces the effective dimensionality per head, leading to suboptimal performance. Based on these observations, we instantiate $\mathrm{TransEnc}(\cdot)$ with one block for Beauty and Yelp, two blocks for Retail Rocket, Gowalla, and Amazon Books, and set the number of attention heads to four in all experiments.

\textbf{\emph{Step Interval Threshold $\delta$.}}
Figure~\ref{fig:params_analysis_append}(a) presents the performance of \emph{BIPCL} under different step interval thresholds $\delta$. On the Beauty dataset, the best performance is achieved at $\delta = 3$, while for the remaining datasets, the optimum is at $\delta = 5$. When $\delta$ is too small, the resulting co-occurrence statistics are sparse and insufficient to capture reliable item relationships. Conversely, overly large thresholds tend to introduce irrelevant co-occurrences, which weaken meaningful structural signals in the graph. Importantly, within a reasonable range of $\delta$, performance varies only mildly, indicating that \emph{BIPCL} is robust to the exact choice of this hyperparameter.

\textbf{\emph{Temperature Parameters $\tau_1$ and $\tau_2$.}}
Following \emph{SimRec}~\cite{simrec}, we set $\tau_1 = 1.0$, corresponding to the standard sampled softmax formulation. Figure~\ref{fig:params_analysis_append}(b) shows that all datasets consistently achieve optimal performance at $\tau_2 = 0.2$. Moving away from this value in either direction results in a substantial drop in performance, highlighting the importance of $\tau_2$ in controlling the sharpness of the contrastive objective and the resulting representation separation. Based on these results, we fix $\tau_2 = 0.2$ for all experiments.

\end{document}